\documentclass[aps,prl,twocolumn,superscriptaddress,10pt,showpacs]{revtex4-1}
\pdfoutput=1

\usepackage{graphicx}
\usepackage{amsmath,amssymb}
\usepackage{hyperref}
\hypersetup{colorlinks=true,
						linkcolor=red,
						anchorcolor=blue,
						citecolor=blue,
						filecolor=blue,
						menucolor=blue,
						urlcolor=blue
}

\usepackage{datetime}
\usepackage{color}
\usepackage[english]{babel}

\begin{document}

\title{Formation and dynamics of anti-ferromagnetic correlations in tunable optical lattices} 

\author{Daniel Greif}
\affiliation{Institute for Quantum Electronics, ETH Zurich, 8093 Zurich, Switzerland}
\affiliation{Department of Physics, Harvard University, Cambridge, Massachusetts 02138, USA}

\author{Gregor Jotzu}
\affiliation{Institute for Quantum Electronics, ETH Zurich, 8093 Zurich, Switzerland}

\author{Michael Messer}
\affiliation{Institute for Quantum Electronics, ETH Zurich, 8093 Zurich, Switzerland}

\author{R\'emi Desbuquois}
\email{desbuquois@phys.ethz.ch}
\affiliation{Institute for Quantum Electronics, ETH Zurich, 8093 Zurich, Switzerland}

\author{Tilman Esslinger}
\affiliation{Institute for Quantum Electronics, ETH Zurich, 8093 Zurich, Switzerland}

\date{\today}

\begin{abstract}
We report on the observation of anti-ferromagnetic correlations of ultracold fermions in a variety of optical lattice geometries that are well described by the Hubbard model, including dimers, 1D chains, ladders, isolated and coupled honeycomb planes, as well as square and cubic lattices. 
The dependence of the strength of spin correlations on the specific geometry is experimentally studied by measuring the correlations along different lattice tunneling links, where a redistribution of correlations between the different lattice links is observed. 
By measuring the correlations in a crossover between distinct geometries, we demonstrate an effective reduction of the dimensionality for our atom numbers and temperatures.
We also investigate the formation and redistribution time of spin correlations by dynamically changing the lattice geometry and studying the time-evolution of the system. 
Timescales ranging from a sudden quench of the lattice geometry to an adiabatic evolution are probed.
\end{abstract}


\maketitle

Understanding the mechanisms underlying quantum magnetism is among the most thought-provoking challenges of quantum many-body physics. 
At the center of these efforts is the interplay between the emergence of magnetic correlations and the underlying lattice geometry \cite{auerbach1994}. 
Extensive research has been carried out using different materials, as well as theoretical and numerical methods, which enhanced the understanding and also triggered unforeseen questions \cite{giamarchi2003, balents2010}. 
A new development is the study of quantum magnetism using ultracold fermionic atoms in optical lattices. 
This technique offers a uniquely direct link between experimental observations and theoretical models, a key element for quantum simulation \cite{Bloch2008d, Esslinger2010c}. 
It also promises unprecedented dynamic control over lattice parameters and geometry \cite{Sebby-Strabley2006a, soltan-panahi2011, tarruell2012, jo2012, Taie2015}, which can give an entirely new perspective on out-of-equilibrium properties of quantum spin systems \cite{Yan2013}. 
Indeed, anti-ferromagnetic spin correlations were recently observed in the Hubbard regime of an optical lattice, first in an anisotropic \cite{greif2013} and later in an isotropic simple cubic geometry \cite{Hart2015}. 
In both situations highly sophisticated computational methods were required for comparison between experiment and theory \cite{Sciolla2013, imriska2014, Hart2015, golubeva2015}. 
Along a different line, making use of ultracold bosons, progress was made in simulating static and dynamic properties of classical and quantum mechanical spin models in theoretically more tractable regimes \cite{simon2011, struck2011, struck2013, fukuhara2013, fukuhara2013b, Hild2014, Brown2015}.

In this letter we explore the emergence of anti-ferromagnetic spin correlations in different lattice geometries of varying dimensionality, also including crossover configurations between different geometries. 	
The dynamic control over the geometries enables us to study the formation dynamics and the redistribution time of spin-correlations, where the explored timescales range from the sudden to the adiabatic regime. 
The starting point of the experiment is a harmonically confined ultracold Fermi gas of $^{40}$K in a balanced two-component spin-mixture with repulsive interactions \cite{greif2013}. 
The atoms are prepared in the two magnetic sub-levels $m_{\mathrm{F}}=-9/2$, $-7/2$ of the $\mathrm{F}=9/2$ hyperfine manifold, and the s-wave scattering length is tuned between $136.4(5)-149.0(3)\,a_0$ via the Feshbach resonance located at $202.1$ G ($a_0$ denotes the Bohr radius). 
For all experiments, the atom number is $140(30) \times 10^3$ with $10$\% systematic error and the temperature is $0.09(1)\,T_{\mathrm{F}}$, where $T_{\mathrm{F}}$ is the Fermi temperature. 
After the preparation, the atoms are loaded into the lowest band of a tunable-geometry optical lattice using an S-shaped ramp lasting $100\,\mathrm{ms}$. 
The lattice consists of several retro-reflected interfering and non-interfering laser beams of wavelength $\lambda=1064\,\mathrm{nm}$, which gives access to a broad variety of lattice geometries \cite{tarruell2012, si}. 
Additionally, in all measurements a 3D harmonic confinement is present in the experiment with a geometric mean trapping frequency of $\bar{\omega}/2\pi = 57(1)\,\mathrm{Hz}$. 

Our experiments are well described by the Fermi-Hubbard Hamiltonian
\begin{equation}
    \hat H   =  -\sum_{\langle ij\rangle,\sigma}t_{ij}(\hat c^\dagger_{i\sigma}\hat c_{j\sigma}+\mathrm{h.c.}) 
    + U\sum_i \hat n_{i\uparrow}\hat n_{i\downarrow} + \sum_{i,\sigma} V_i \hat n_{i\sigma}\;,
\end{equation}
with tunneling $t_{ij}$ between nearest neighbours $\langle ij\rangle$ and repulsive on-site interaction $U$. 
Here $\hat c^\dagger_{i\sigma}$ and $\hat c_{i\sigma}$ denote the fermionic creation and annihilation operators for the two spin states $\sigma\in\{\uparrow,\downarrow\}$, the density operator is denoted by $\hat n_{i\sigma}=\hat c^\dagger_{i\sigma} \hat c_{i\sigma}$ and $V_i$ is the trap energy. 
The different lattice geometries are realized experimentally by independently adjusting the specific values of $t_{ij}$ for each of the six nearest neighbour links per lattice site of an underlying simple cubic lattice. 
Their strength is controlled via the power of the lattice laser beams \cite{si}. 
In all measurements presented in the following, we adjust the scattering length with a Feshbach resonance such that $U/h=0.87(2)\,\mathrm{kHz}$ and set the total bandwidth for non-interacting particles to $W/h=2.6(1)\,\mathrm{kHz}$ \footnote{All Hubbard parameters are calculated from the lattice potential using Wannier functions, which are obtained using band-projected position operators \cite{uehlinger2013}.}.

After loading the atoms into the desired lattice geometry, we measure the trap-averaged magnetic spin correlations emerging on neighbouring sites in the low-temperature many-body state of the quantum gas. 
Our detection is similar to the method used in previous experiments \cite{greif2013}, and is presented in full detail in the Supplemental Information \cite{si}. 
The spin correlations are measured on every second lattice link, between nearest neighbours $i$ and $i+1$, and along the transverse spin axis
\begin{equation}
C_{i,i+1} = -\langle\hat{S}^x_i \hat{S}^x_{i+1}\rangle- \langle\hat{S}^y_i \hat{S}^y_{i+1}\rangle.
\end{equation}
Here $\hat{S}^{x,y,z}_i$ denote the standard spin vector operators for a spin-1/2 system on site $i$, and $\langle...\rangle$ denotes the trap average. 
For SU(2) symmetry, $C_{i,i+1}$ is equal to $-2\langle \hat{S}^z_i\hat{S}^z_{i+1}\rangle$. 
The detection protocol allows us to measure both anti-ferromagnetic and ferromagnetic configurations, corresponding to positive and negative values of $C_{i,i+1}$, respectively. 

\begin{figure}[tb]
    \includegraphics{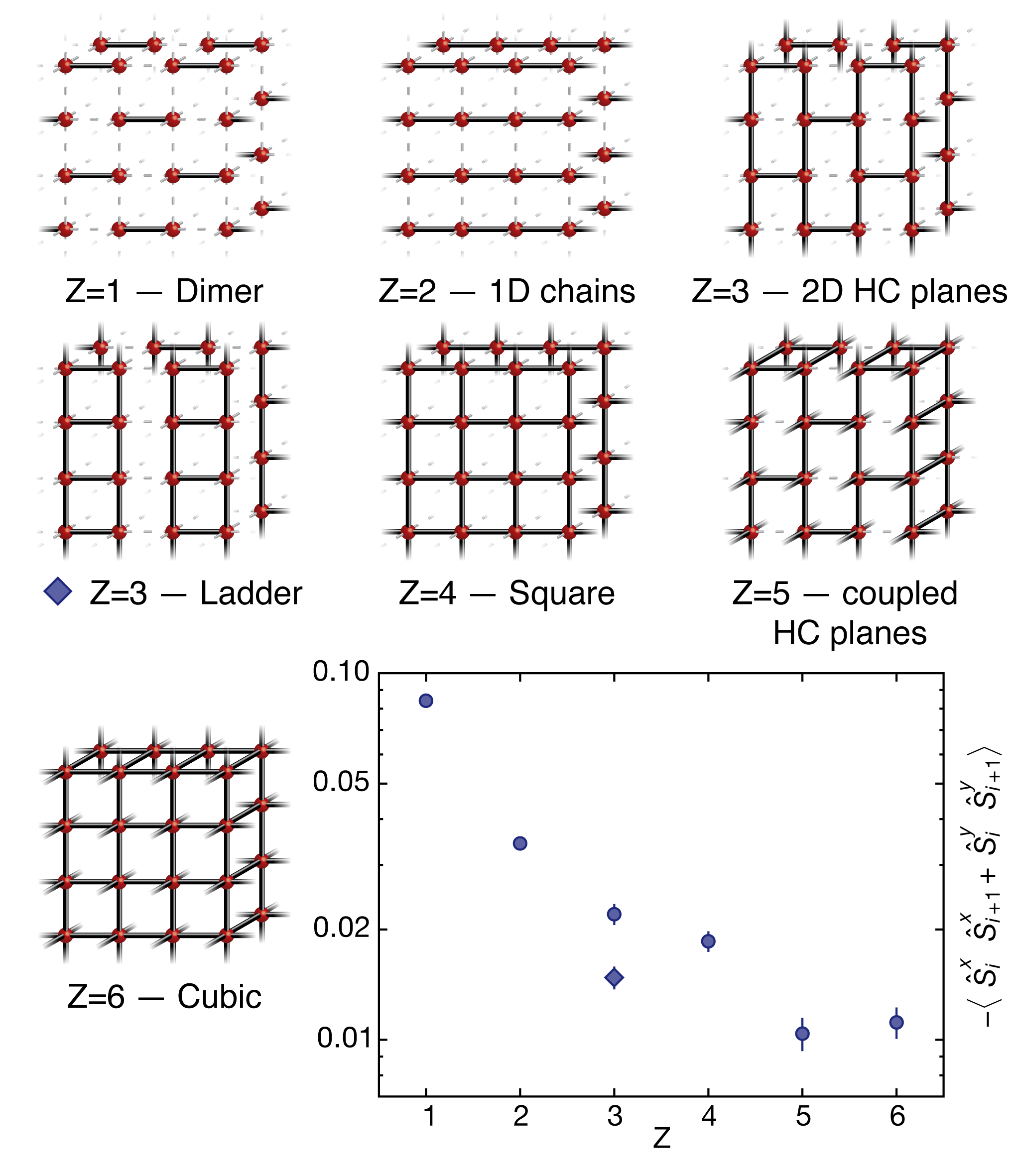}
    \caption{Experimental observation of the dependence of anti-ferromagnetic correlations on lattice geometry. The trap averaged correlator $-\langle\hat{S}^x_i \hat{S}^x_{i+1}\rangle- \langle\hat{S}^y_i \hat{S}^y_{i+1}\rangle$ is measured along the strong links for various lattice geometries, which differ in the number of strong nearest neighbour links $Z$. A schematic view of the lattice geometries is additionally shown, where the strong tunneling links are indicated by a bar. For all data points the bandwidth $W/h=2.6(1)\,\mathrm{kHz}$ and the on-site interaction $U/h=0.87(2)\,\mathrm{kHz}$ is constant. Error bars denote the standard error of $50$ measurements.}
	\label{fig1}
\end{figure}

In a first measurement, we investigate the strength of spin correlations in several different lattice geometries. 
Starting from an underlying simple cubic lattice, the tunneling is enhanced along $Z$ nearest neighbour links and takes the value $t_s$, whereas the tunneling along the remaining $6-Z$ links is $t_s/5$. 
The geometries realized in this manner are, sorted by increasing number of strong nearest neighbour links $Z$: dimerized, 1D chains, honeycomb planes, ladders, square, coupled honeycomb planes, and cubic (see Fig. 1).
We measure the correlations along the strong tunneling links.
As shown in Fig. 1, the strength of the correlations depends on the specific geometry with values ranging between $0.084(1)$ and $0.010(1)$ for the trap averaged value, and is generally smaller for a larger number of strong tunneling links. 
In the isotropic cubic lattice as well, which has the largest value of Z (Z=6), we detect anti-ferromagnetic correlations in the system \cite{Hart2015}. 

The observed dependence of the spin correlator on $Z$ can be understood with two simple arguments in a homogeneous system.
Owing to the isolated nature of the system, the total entropy rather than the temperature is constant for different $Z$. 
First, with a finite entropy, the presence of two different energy scales associated with different tunnelings directly affects the magnetic correlations and leads to a redistribution of spin correlations between the strong and the weak links.
For a large number of weaker tunneling links, more low-energy states are accessible. 
Thus, a finite entropy mainly leads to thermal fluctuations within theses states, and the magnetic correlator on the strong links is high. 
However, if the number of weak links decreases, thermal fluctuations along the weak links alone are not sufficient to account for the total entropy, and additional thermal fluctuations are also distributed on the strong links, therefore reducing the strong link correlator.
Second, even at zero entropy, quantum fluctuations play a significant role. In lower dimensions this generally leads to enhanced short-range spin correlations \cite{Gorelik2012}.
In both cases, the correlator is expected to decrease as $Z$ is increased.
This is in accordance with previous measurements in the specific cases of a dimerized lattices ($Z=1$) and 1D chains ($Z=2$) \cite{greif2013, imriska2014}. 

While these two effects predict a dependence only on $Z$, the lattice geometry itself (for the same value of $Z$) will also affect the strength of the spin correlations, most importantly at low temperatures. 
In fact, in the limit of vanishing temperatures the state of the system and its phase diagram will be entirely determined by the interplay between geometry and magnetic ordering. 
The different values observed for the two $Z=3$ geometries (ladder and honeycomb planes) suggest that effects of the lattice geometry are already starting to play a role at the temperatures reached in the experiment. 
An analysis of the exact extent of this effect may be achieved by a detailed experiment-theory comparison, which is beyond the scope of this work.

In addition to the aforementioned contributions, the underlying harmonic confinement also plays a central role for the value of the trap-averaged correlator. 
Within the local density approximation, both the chemical potential at the center of the trap and the temperature are determined by the total atom number and entropy. 
As the equation of state of the system depends on the lattice geometry and $Z$, both the density and entropy distribution change with the geometry, which directly affects the magnetic correlator.

\begin{figure}[tb]
    \includegraphics{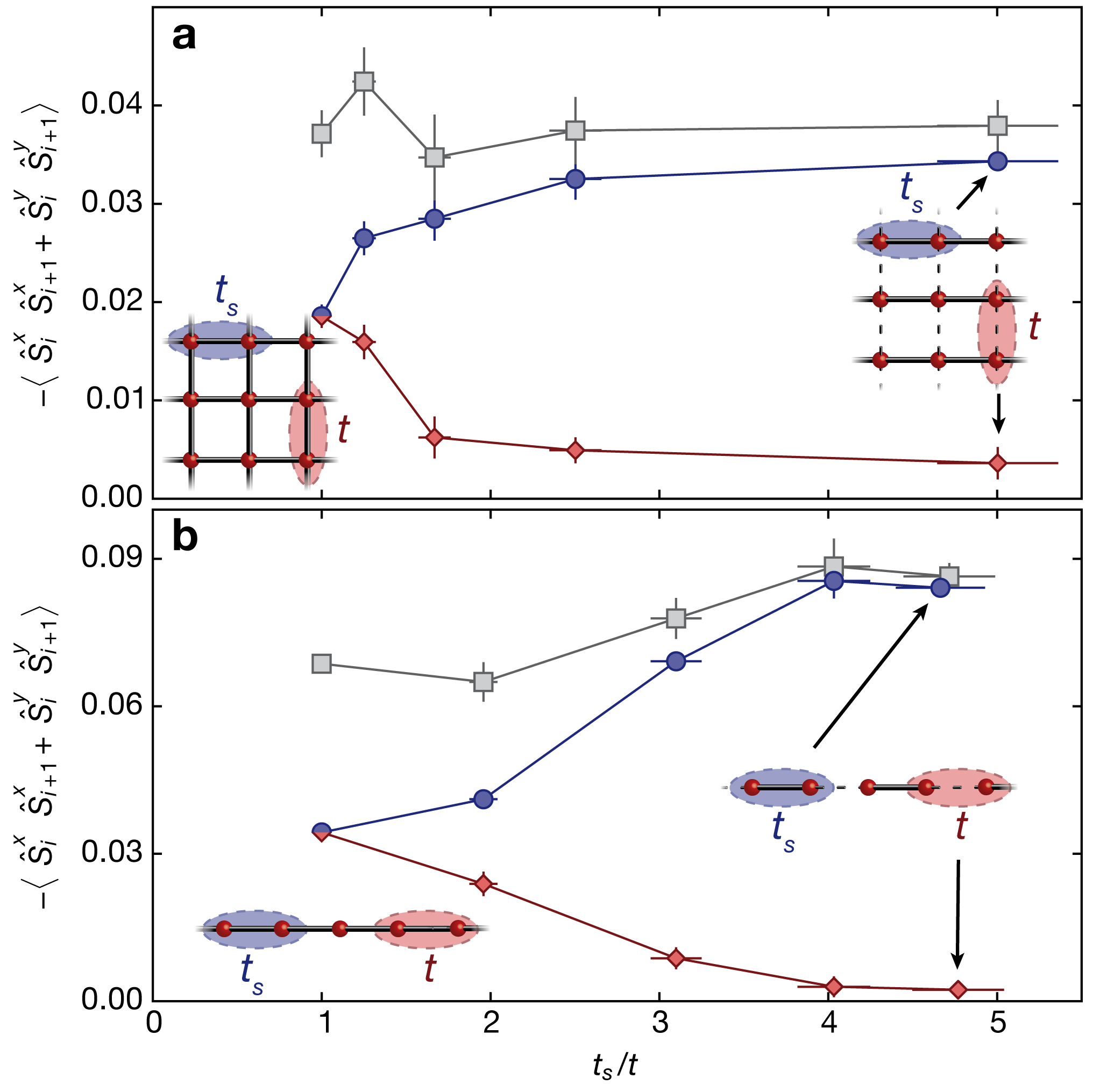}
    \caption{Spin correlations for a crossover between two different lattice geometries. (a) Scan between a square ($Z=4$) and a 1D chain ($Z=2$) geometry and (b) scan between a 1D chain ($Z=2$) and a dimerized ($Z=1$) geometry. In both cases, the strong to weak tunneling ratio $t_s/t$ varies between $1$ and $5$. Blue and red symbols denote the measured spin correlations along the strong and weak links respectively. Gray symbols show their sum. The error bar on the tunneling ratio denotes the uncertainty on the lattice parameters, while the data is the mean $\pm$ the standard error of at least 25 measurements.}
	\label{fig2}
\end{figure}

To further study the impact of geometry on magnetic correlations, we measure their strength for a crossover regime between two lattice geometries with different numbers of strong links $Z$. 
In the experiment geometries with a different $Z$ can be smoothly connected by adjusting the strength of the individual tunneling links \cite{si}. 
We scan between a square ($Z=4$) and 1D chain ($Z=2$) geometry by choosing the strong-to-weak tunneling ratio $t_s/t$ in the range $1-5$, while keeping the tunneling in the orthogonal direction at $t_s/5$, see Fig. 2a. 
A second scan between a 1D chain and dimerized ($Z=1$) geometry is shown in Fig. 2b.
We measure spin correlations either along the strong or the weak tunneling links.
In both cases, correlations on each link start from the same value. 
As $t_s/t$ is increased, the correlations along the strong links are enhanced, whereas the correlations along the weak links decrease. 
Interestingly, the correlations change more rapidly with increasing $t_s/t$ for the scan in Fig. 2a as compared to Fig. 2b, which is a consequence of the underlying lattice geometry. 
For the final configuration $t_s/t=5$, the correlations on the weak link have nearly vanished, whereas the correlations on the strong links have saturated at a high value. 
This indicates that the thermal fluctuations occur predominantly on the weak links. 
Consequently, the weaker couplings can be neglected in this case, and the dimensionality of the lattice is effectively reduced for our total entropy and atom number. 
Interestingly, the reduction of the dimensionality occurs at different ratios of $t_s/t$, depending on which geometry is considered.

These measurements demonstrate that spin correlations redistribute between the strong and weak links when changing $Z$. 
Yet, this does not necessarily imply that the sum of spin correlations is constant. 
We find the sum of correlations to be approximately constant in the scan of Fig. 2a, whereas it increases significantly with dimerization in the scan of Fig. 2b. 
This observation might be related to the opening of a finite energy gap in the energy spectrum for a strongly dimerized lattice, which causes entropy redistribution within the trapped system and enhances the overall spin correlation strength \cite{greif2013}. 

\begin{figure*}[tb]
		\centering
    \includegraphics[width=\textwidth]{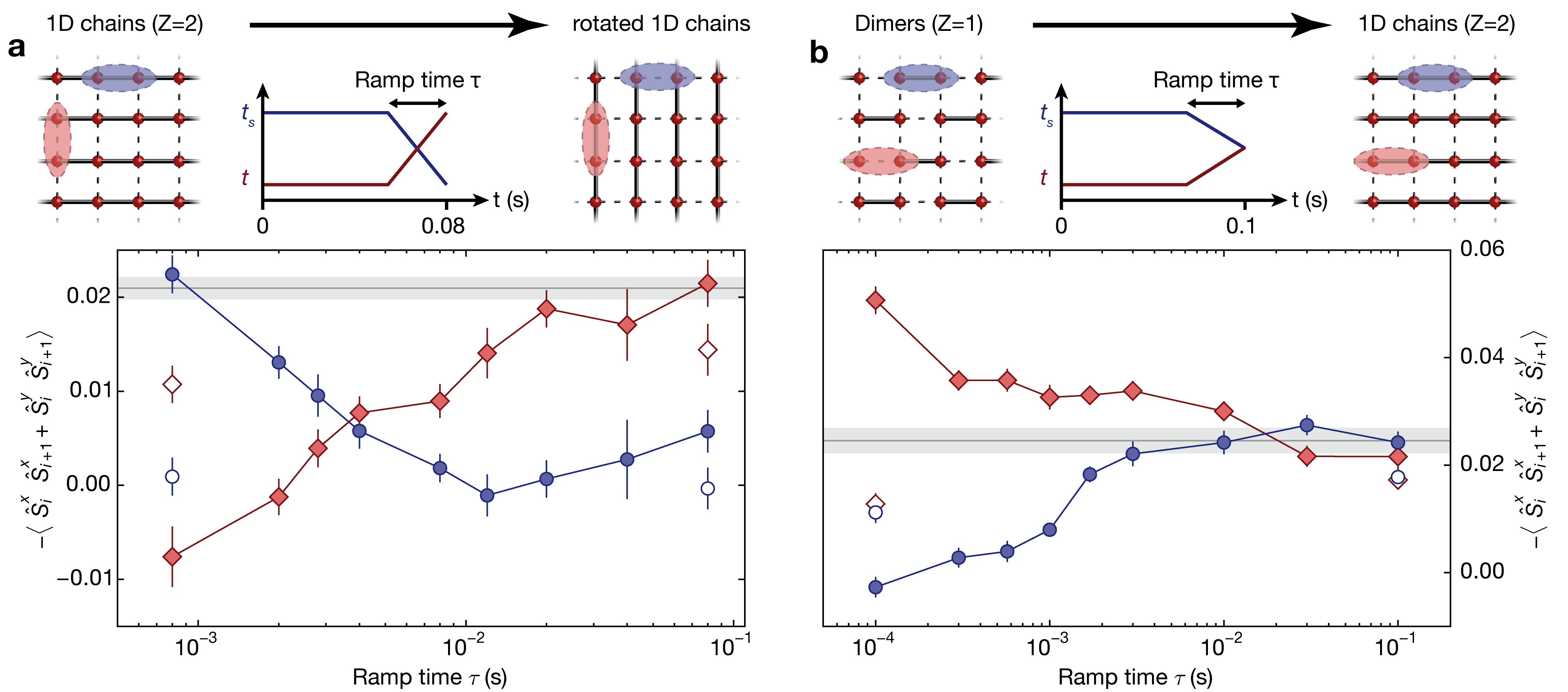}
    \caption{Dynamics of spin correlations. The lattice is ramped within a time $\tau$ {\bf a} from a 1D chain geometry to the same 1D chain geometry rotated by $90^{\circ}$ and {\bf b} from a dimerized to 1D chain geometry \cite{si}. Blue and red points denote the measured spin correlations along the previously and new strong links. For all closed symbols the correlations are measured immediately after the ramp, whereas open symbols include an additional wait time of $50\,\mathrm{ms}$. The gray solid lines indicate the reference value for the correlations in the initial geometry without lattice ramp, and the shading denotes the error on this value. Error bars are the standard error of at least 25 measurements.}
	\label{fig3}
\end{figure*}

The tunability of our lattice also allows us to experimentally measure the timescales for the formation and redistribution of spin correlations when dynamically changing the lattice geometry. 
This is done by measuring the strength of spin correlations when the lattice geometry is changed on variable timescales. 
For simplicity we start with a ramp where the initial and final lattice geometry are the same up to a rotation: starting from a 1D chain geometry we ramp via a square lattice to a 1D chain lattice again, but with strong tunneling along the perpendicular direction. 
We always include an additional wait time before the ramp such that the total time in the optical lattice is constant, see Fig. 3a. 
The spin correlations are measured immediately after the ramp along the two different directions. 

The observed dependence of the spin correlations on the total ramp time $\tau$ is shown in Fig. 3a. 
For fast ramps, $\tau<1\mathrm{ms}$, the spin correlations remain unchanged, at the values without ramping. 
Here, a non-equilibrium state is formed with several charge and spin excitations, which decay when allowing for an additional wait time after the ramp - thus changing the detected value of spin correlations (open symbols in plot Fig. 3a). 
On intermediate ramp times, $\tau\sim4\,\mathrm{ms}$, the spin correlations change symmetrically along the two directions. 
This timescale is comparable to the underlying tunneling time between $h/t_s=2\,\mathrm{ms}$ and $h/t=10\,\mathrm{ms}$ during the ramp. 
For very slow ramp times we observe within error bars a $100$\% transfer of spin correlations from the previously strong to the new strong links. 
When waiting $100\,\mathrm{ms}$ in this case, the magnetic correlations decrease (most likely due to underlying heating of the gas), but agree with the value when loading directly from the harmonic trap to the final lattice geometry and waiting for the same total time. 
These observations are in agreement with a fully adiabatic ramp to an equilibrated final state for the slowest ramp times, as the initial and final lattice geometries are the same with the same density and entropy distributions. 

The situation changes considerably for a ramp with different initial and final lattice geometries. 
Here, we start from a dimerized lattice with a ratio of $t_s/t=5$ for adjacent tunneling links and ramp to a 1D geometry without dimerization. 
Immediately after the lattice ramp we measure the spin correlations on the initially strong and weak links along the 1D direction, see Fig. 3b. 
As before, for fast ramps the spin correlations cannot redistribute and are nearly unchanged as compared to the case without ramp. 
When adding an additional wait time after the fastest ramp, the spin correlations change, signaling a decay of the created excitations in this case. 
The behavior is different for intermediate ramp times: while the correlations on the initially strong links decrease very quickly, slower ramp times $\tau\sim 1\,\mathrm{ms}$ are necessary for the correlations on the initially weak links to change. 
This may originate from the difference in the overall tunneling timescale during the ramp for the two links. 
For the slowest ramps, the correlators along the original strong and weak link are identical.
With an additional wait time, they both decay to the same value, again owing to heating.
This indicates a final state close to equilibrium. 
In contrast to the previous measurement, the gap between the ground and excited states closes during the ramp, since the singlet-triplet gap has vanished in the non-dimerized geometry. 
Yet, the observed spin correlation value agrees with a reference when loading directly into the final lattice geometry and holding the remaining time. 
Consequently, ramp times corresponding to a few tunneling times are already sufficient to reach equilibrium. 

In conclusion, we have observed anti-ferromagnetic correlations in a variety of lattice geometries and studied the redistribution of correlations between strong and weak links. 
Extending our work to lower temperatures will allow addressing open questions on the low-temperature phase diagram of the Hubbard model in complex lattice geometries. 
In combination with recent efforts on creating spin-dependent optical lattices with very low underlying heating \cite{Jotzu2015}, the anisotropic XXZ Heisenberg model and the Falikov-Kimball model could be realized and studied experimentally with ultracold fermions \cite{Freericks2003}. 
Our results on quantum spin dynamics demonstrate that ultracold fermions in optical lattices are well suited to study open questions in out-of-equilibrium many-body spin systems, where theoretical methods become extraordinarily difficult \cite{Polkovnikov2011a, Eisert2015}.
Furthermore, the observed rapid formation timescales of spin correlations offer very promising perspectives for the implementation of sophisticated entropy redistribution schemes based on trap shaping and dynamically changing lattice geometries, which are expected to result in overall lower temperatures \cite{Ho2009a, bernier2009, lubasch2011}. 

\begin{acknowledgments}
We would like to thank Frederik G\"org, Lei Wang and Jakub Imri\v ska for insightful discussions and valuable contributions. We acknowledge SNF, NCCR-QSIT, QUIC (Swiss State Secretary for Education, Research and Innovation contract number 15.0019) and SQMS (ERC advanced grant) for funding. R.D. acknowledges support from ETH Zurich Postodoctoral Program and Marie Curie Actions for People COFUND program.
\end{acknowledgments}

\clearpage

\makeatletter
\setcounter{section}{0}
\setcounter{subsection}{0}
\setcounter{figure}{0}
\setcounter{equation}{0}
\setcounter{table}{0}
\renewcommand{\bibnumfmt}[1]{[S#1]}
\renewcommand{\citenumfont}[1]{S#1}
\renewcommand{\thefigure}{S\@arabic\c@figure}
\renewcommand{\theequation}{S\@arabic\c@equation}
\renewcommand{\thetable}{S\@arabic\c@table}

\section{Supplemental material}
   
\subsection{Preparation and optical lattice}

For our measurements we prepare a quantum degenerate cloud of $^{40}\, \mathrm{K}$ fermions, harmonically confined in a crossed optical dipole trap. 
This is done by evaporatively cooling a balanced spin mixture of the two magnetic sub-levels $m_{\mathrm{F}}=-9/2$, $-7/2$ of the $\mathrm{F}=9/2$ hyperfine manifold. 
We reach temperatures of $0.09(1)\,T_{\mathrm{F}}$, where $T_{\mathrm{F}}$ is the Fermi temperature.
This corresponds to an entropy of $0.87 k_B$ per particle in the harmonic trap. 
The atom number is set to a value of  $140(30) \times 10^3$ for all measurements.
The repulsive on-site interaction energy $U$ is tuned via a Feshbach resonance located at $202.1\, \mathrm{G}$ by changing the s-wave scattering length between $a = [136.4(5)-149.0(3)]\,a_0$, where $a_0$ denotes the Bohr radius.  

After the preparation, the fermions are loaded into a tunable-geometry optical lattice within $100\,\mathrm{ms}$ using an S-shaped ramp. 
The lattice potential is created by several retro-reflected interfering and non-interfering laser beams of wavelength $\lambda=1064\,\mathrm{nm}$, which gives access to a broad variety of lattice geometries described by the following potential \cite{Tarruell2012}:  
\begin{eqnarray} V(x,y,z) & = & -V_{\overline{X}}\cos^2(k
x+\theta/2)-V_{X} \cos^2(k
x)\nonumber\\
&&-V_{Y} \cos^2(k y) -V_{\widetilde{Z}} \cos^2(k z) \nonumber\\
&&-2\alpha \sqrt{V_{X}V_{Y}}\cos(k x)\cos(ky)\cos\varphi , \label{eqlattice}
\end{eqnarray}
where $V_{\overline{X},X,Y,\widetilde{Z}}$ are the lattice depths of each single beam in direction $x,y,z$ given in units of the recoil energy $E_R=h^2/2m\lambda^2$, $k=2\pi/\lambda$, visibility $\alpha=0.81(1)$ and $h$ denotes the Planck constant. 
We set $\theta=\pi$ and interferometrically stabilize the phase $\varphi=0.00(3)\pi$.
The lattice depths $V_{\overline{X},X,Y,\widetilde{Z}}$ are independently calibrated using Raman–Nath diffraction on a $^{87}\mathrm{Rb}$ Bose–Einstein condensate.

Each lattice geometry shown in Fig. 1 is obtained by individually adjusting the lattice depths $V_{\overline{X},X,Y,\widetilde{Z}}$.
Following this procedure the connectivity $Z$ (i.e. the number of strong nearest neighbor links) can be varied from 1 (dimerized lattice) to 6 (cubic lattice). 
This is achieved by enhancing the tunneling rate of $Z$ nearest-neighbor lattice bonds, with respect to the remaining $6-Z$ bonds per site. 
Due to the tunability of our optical lattice we can realize the seven individual lattice geometries shown in Fig. 1, as well as the crossover configurations, such as the ones shown in Fig. 2. 
Additionally it allows us to ramp from one lattice geometry to another on a variably time-scale, which is used for the dynamics measurements shown in Fig. 3. 

To set the correct lattice depths we calculate all Hubbard parameters numerically from the lattice potential using Wannier functions.
Our computational method is based on their definition as eigenstates of band-projected position operators \cite{Uehlinger2013}. 
From the overlap integrals of the Wannier functions we obtain the tunneling $t_{ij}$ between nearest neighbors and the interaction energy $U$. 
For all our measurements, we choose the lattice depths in such a way that the total bandwidth of the single orbital tight-binding model is always $W/h = 2.60(6)\, \mathrm{kHz}$, independent of the particular lattice geometry. 
The expression is given by $W=2\sum_i t_{0i}$, where $i$ denotes the sum over all 6 nearest neighbors of any given lattice site in the underlying simple cubic lattice. 
Since the Wannier functions depend on the exact geometry and lattice depths, we have to adjust the scattering length by tuning the magnetic field of the Feshbach resonance to ensure a constant on-site interaction $U/h=867(15)\, \mathrm{Hz}$. The exact parameters for all the lattice geometries with different connectivities realized in the experiment are summarized in Table \ref{table_lattice_configurations}. 

Due to the changing harmonic confinement as a result of the intensity variation of the lattice beams, we have to adjust the optical power of the additional dipole trap to set a constant geometric mean trapping frequency of $\bar{\omega}/2\pi = 57(1)\,\mathrm{Hz}$ for all measurements.
However, the individual trapping frequencies in $x$-, $y$- and $z$-direction of the harmonic confinement are changing for each lattice geometry.

\subsection{Detection scheme for spin-correlations}

\begin{figure*}[t]
    \includegraphics{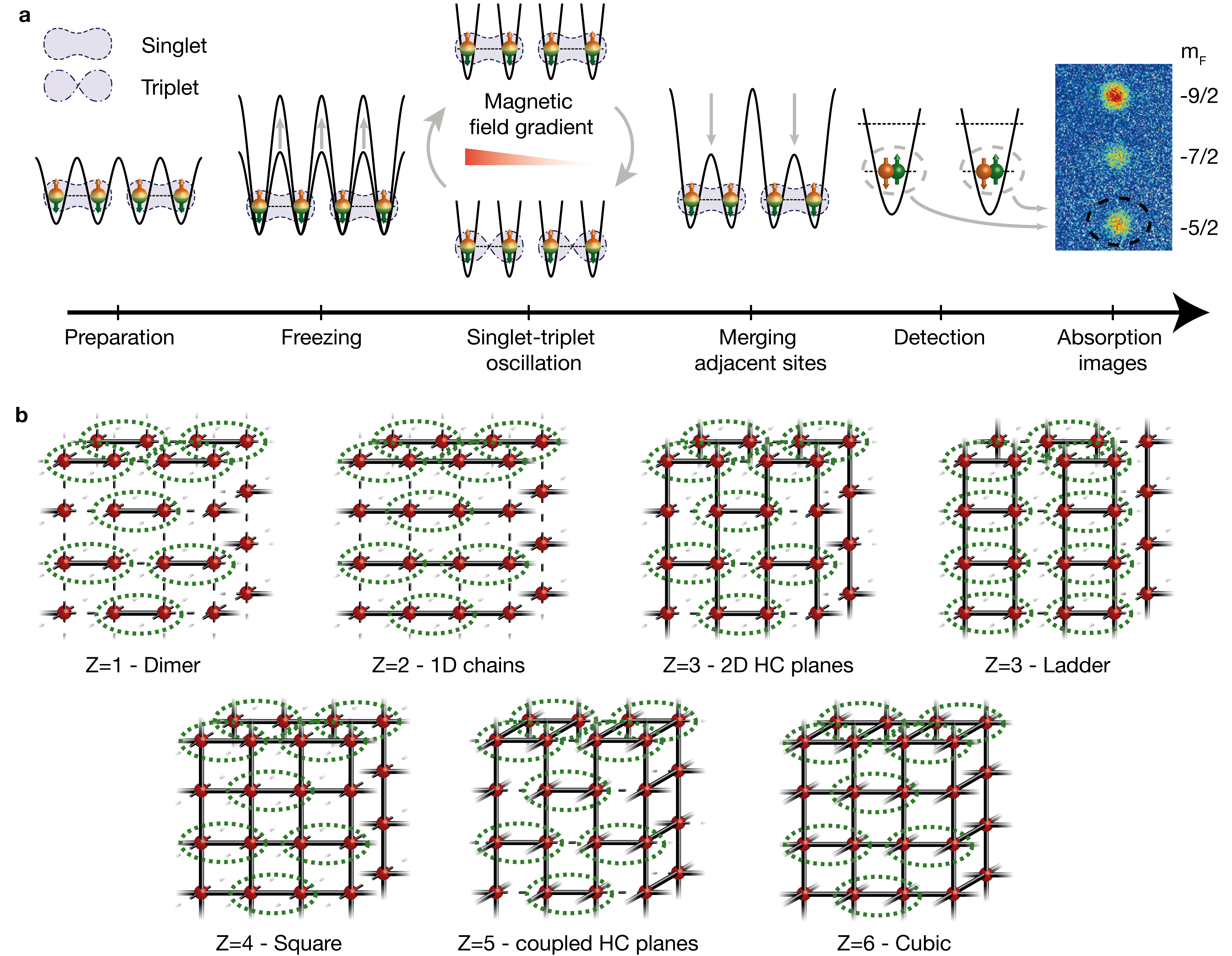}
    \caption{Probing nearest neighbor spin correlations.    
    (a) The detection scheme of the anti-ferromagnetic correlations contains several steps and follows closely the procedure used in previous experiments \cite{Greif2013}. 
    After preparing the system in a thermalized many-body state with varying geometry, we freeze and convert it into a deep simple cubic lattice. 
    In a second step we apply a magnetic field gradient which causes coherent oscillations between the singlet and the triplet state on neighboring sites. 
    In a final step we remove the magnetic field gradient and merge pairs of neighboring sites. 
    Depending on the oscillation time we detect either the number of singlets or the number of triplets corresponding to the initial many-body state.     
    (b) An overview of the set of links between nearest neighbors, along which the spin correlations are measured for the various geometries in Fig. 1 of the main manuscript, are marked in colored ellipsoids (note that the ladder geometry has been rotated for clarity).
   	}\label{detection_scheme}
\end{figure*}

The detection scheme for nearest-neighbor anti-ferromagnetic spin correlations consists of several steps (see Fig. \ref{detection_scheme}) and closely follows the procedure used in previous experiments \cite{Greif2013}. 
After preparing the many-body state in the chosen lattice geometry, we perform a sudden ramp of the lattice depth to freeze out the atomic motion. 
For this we always ramp into a deep simple cubic lattice with $V_{\overline{X},X,Y,\widetilde{Z}}=[30,0,30,35]\,E_R$  within $100\,\mathrm{\mu s}$ independent of the starting geometry of the optical lattice (i.e. Z=1..6).
The overall measurement projects the state of the initial lattice onto the singlet  $\left|s\right\rangle=(\left|\uparrow,\downarrow\right\rangle-\left|\downarrow,\uparrow\right\rangle)/\sqrt{2}$ and triplet states $\left|t_0\right\rangle=(\left|\uparrow,\downarrow\right\rangle+\left|\downarrow,\uparrow\right\rangle)/\sqrt{2}$ on the respective set of links where the spin correlations are measured (see Fig. \ref{detection_scheme}b).
From this we determine the fraction of atoms forming singlets $p_s$ and triplets $p_{t_0}$ on neighboring sites. 
An imbalance in these two fractions corresponds to magnetic ordering on neighboring sites. 
The population imbalance is proportional to the transverse spin correlator \cite{Greif2013}
\begin{equation}
C_{i,i+1} = -\langle\hat{S}^x_i \hat{S}^x_{i+1}\rangle- \langle\hat{S}^y_i \hat{S}^y_{i+1}\rangle = (p_s - p_{t_0})/2.
\end{equation}
which is measured on every second lattice link with nearest neighbors $i$ and $i+1$ and along the transverse spin axis (see Fig. \ref{detection_scheme}b).
Due to our detection process all quantities are averaged over the entire atomic cloud. 
As doubly occupied sites do not contribute to the spin correlator and hinder the detection process, they are removed in the deep cubic lattice using a series of Landau-Zener transfers before detecting the singlet and triplet fractions. 

We measure the singlet and triplet fractions by subsequently applying a magnetic field gradient, which lifts the energy degeneracy for atoms  with opposite spins on neighboring sites. 
This induces a coherent oscillation between the singlet and the triplet states. 
The singlet-triplet oscillations (STO) have a frequency of $\nu = \Delta_{\mathrm{STO}}/h$, where $\Delta_{\mathrm{STO}}$ is the energy splitting, and are only visible if the initial amount of singlets and triplets is different, as their respective time evolution is exactly out of phase. After a variable oscillation time we remove the magnetic field gradient and merge pairs of adjacent sites into a single site. 
We use a $10\,\mathrm{ms}$ linear ramp from the deep cubic lattice into a deep checkerboard lattice ($V_{\overline{X},X,Y,\widetilde{Z}}=[0,25,30,35]\,E_R$), where tunneling is still suppressed and the number of sites is divided by two. 
During merging, the $\left|s\right\rangle$ and $\left|t_0\right\rangle$ states are mapped onto different bands due to their distinct symmetry of the two-particle wave function. 
The (spatially symmetric) singlet state is mapped to two atoms in the lowest band of the final state, while the (spatially anti-symmetric) triplet state evolves into a final state with one atom in the first excited band and one atom in the lowest band. 
By adjusting the oscillation time to a maximum or minimum of the STO we can then detect the fraction of atoms in the singlet and triplet state by measuring the fraction of atoms on doubly occupied sites with both atoms in the lowest band. 
This is accomplished using interaction dependent rf-spectroscopy, which transfers the $m_F = -7/2$ atoms on doubly occupied sites in the lowest band to the initially unpopulated $m_F = -5/2$ state \cite{Joerdens2008}.

After all these steps we measure the atomic populations in the different spin states. We ramp down the optical lattice potential and dipole trap within 20ms and apply a magnetic field gradient to separate the different $m_F$ states in a Stern-Gerlach measurement during ballistic expansion. 
Finally we take an absorption image and apply Gaussian fits to the density distribution of each spin component to determine the number of atoms in each spin state (exemplary image shown in Fig. \ref{detection_scheme}).
For the double-occupancy measurement we take into account the independently calibrated detection efficiency of 91(3)\% and the background offset in double occupancy of 2.2\% is subtracted (the latter however does not change the measured value of the spin correlator). In addition, we take into account the separately calibrated decay of the singlet-triplet oscillations during the entire detection sequence \cite{Greif2013}.

\subsection{Calibration of singlet-triplet oscillations}

\begin{figure}
    \includegraphics{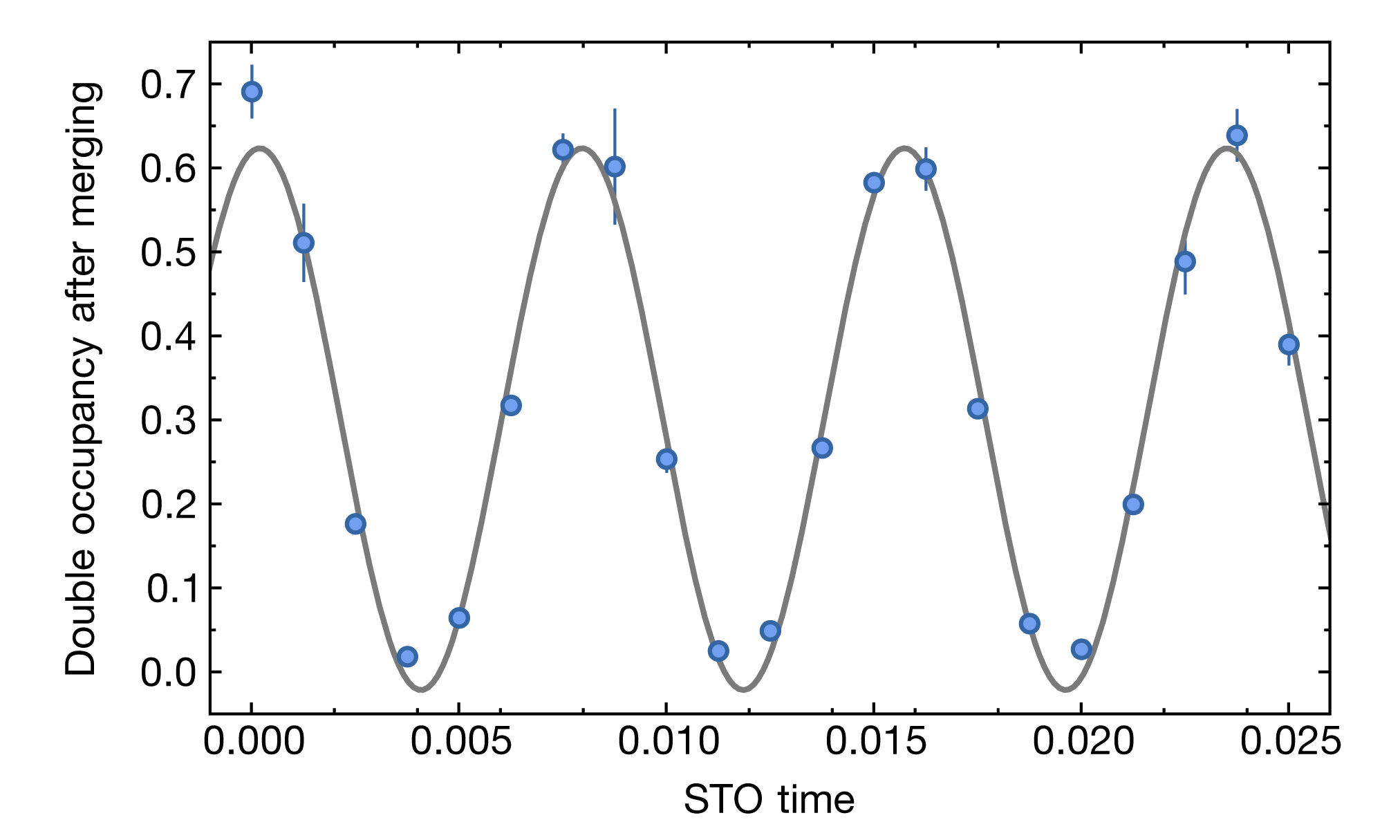}
    \caption{Full singlet-triplet oscillation. 
    We measure the double occupancy after merging as a function of oscillation time during a singlet-triplet oscillation, which is induced by a magnetic field gradient in the deep cubic lattice.
    For this calibration a far-from-equilibrium initial state was prepared with a large number of singlets. 
    The observed phase shift is partially caused by a residual slow STO (with a period $>100 \, \mathrm{ms}$) occurring in the deep cubic lattice already before applying the magnetic field gradient (originating from residual magnetic gradients). 
    Error bars show the standard deviation.
    }\label{fig_fullSTO}
\end{figure}

To test and calibrate the performance of the detection scheme for measuring spin correlations using singlet-triplet oscillations, we prepare an initial state with a large number of singlets and very small number of triplets. 
We load an attractive spin-mixture at $a=-770(50)\,a_0$ into a deep checkerboard lattice with $V_{\overline{X},X,Y,\widetilde{Z}}=[0, 3, 3, 7]\,E_R$, which leads to a large number of doubly occupied sites. 
We first freeze this geometry and then ramp to repulsive interactions with $a=106.5(9)\,a_0$.
In a second step we split the sites of the checkerboard lattice by ramping to the same deep cubic lattice which is used in the first step of the detection scheme. 
This converts the double occupied sites to singlets. 
Although this initial state may be far from equilibrium owing to its preparation, it is well suited for calibration purposes as it contains a very large imbalance in the singlet and triplet fractions, and hence a large value for the nearest-neighbor spin correlator. 
After the initial preparation, we apply the entire detection protocol and measure the double occupancy after merging as a function of oscillation time with magnetic field gradient, see Fig. \ref{fig_fullSTO}.
The fraction of atoms forming singlets (triplets) is then given by the maximum (minimum) value of the measured double occupancy of the STO, which is obtained from a sinusoidal fit to the data.  
We use this measurement to calibrate the magnetic field gradient and determine the singlet-triplet oscillation frequency. 
The actual phase of the STO, which originates from residual magnetic gradients, is determined by preparing a thermal many-body state in the dimerized geometry of the measurement sequence.
From this we determine the maximum and minimum oscillation time, which is used for all measurements in the main manuscript. 

\subsection{Fig 1: Connectivity scan}

Using the tunable geometry optical lattice we can individually address and adjust the specific tunneling values $t_{ij}$ for all six links connecting one lattice site.
To measure the dependence of the correlations on the geometry we enhance the tunneling $t_s$ on $Z$ nearest neighbor links of the underlying simple cubic lattice with Z ranging between 1 and 6, while the tunneling on the remaining links is set to a much smaller value of $t_s/5$.
If all tunneling links are set to the same value then $t_s/h=217(10)\,\mathrm{Hz}$ and we realize a simple cubic lattice.
Table \ref{table_lattice_configurations} summarizes the parameters of each of the seven lattice geometries realized for the measurements in Fig. 1. 
The fraction of singlets and triplets is measured on a set of strong links for each data point, see Fig. \ref{detection_scheme}b. 
In the experiment (see Fig. 1) we choose for each geometry to always merge the same strong links.
For the 1D chain, square and cubic lattice we independently verified that the signal strength of the spin correlator does not depend on the exact choice of the merged strong link.

\begin{figure}
    \includegraphics{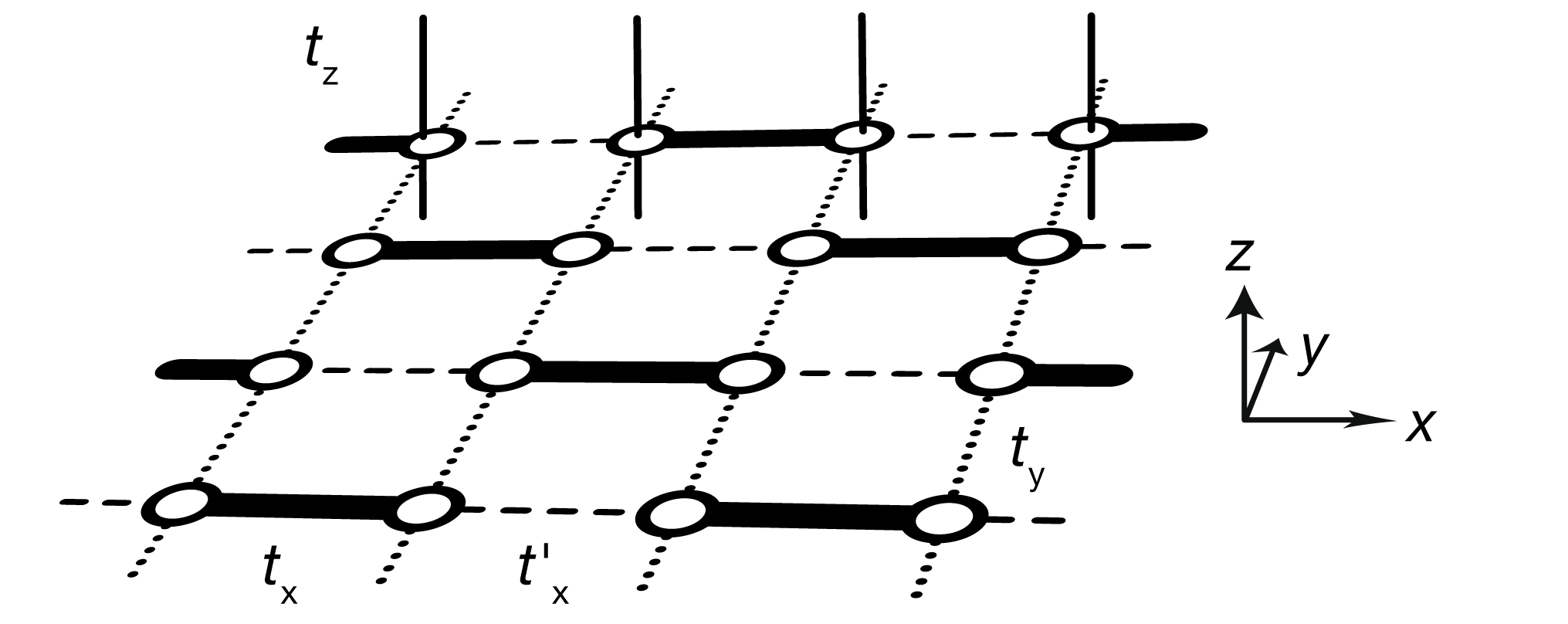}
    \caption{Overview of the different tunneling links in the tunable-geometry optical lattice, which are all independently tunable.
    }\label{fig_lattice_schematics}
\end{figure} 

\begin{table*}[!t]
\begin{center} 
\begin{tabular}{ | l | c | c | c | c | c | c | c | c | c | } 
\hline Lattice & \,Z\, & $V_{\overline{X},X,Y,\widetilde{Z}}\,[E_R]$ & $a \, [a_0]$ & $U/h \, [\mathrm{Hz}]$ & $t_{\mathrm{x}}/h \, [\mathrm{Hz}]$ & $t_{\mathrm{x}}'/h \, [\mathrm{Hz}]$ & $t_{\mathrm{y}}/h \, [\mathrm{Hz}]$ & $t_z/h \, [\mathrm{Hz}]$ & $W/h \, [\mathrm{kHz}]$
\\ \hline  Dimer & 1 & [4.925, 0.073, 7.9, 8.19] & 136.4(5) & 866(15) & 635(30) &136(4) & 130(8) & 130(7) & 2.60(8)
\\  1D & 2 & [3.205, 0, 9.6, 9.6] & 139.1(5) & 867(15) & 464(12) & 464(12) & 93(6) & 93(6) & 2.60(4)
\\  2DHC & 3 & [7.545, 0.258, 3.93, 10.7] & 137.5(5) & 868(15) & 352(24) & 75(3) & 363(12) & 72(5) & 2.60(8)
\\  Ladder & 3 & [7.345, 0.081, 10.45, 4.16] & 136.9(5) & 867(15) & 354(23) & 73(3) & 73(5) & 361(12) & 2.60(6)
\\  2D & 4 & [4.93, 0, 4.93, 11.6] & 140.9(4) & 867(15) & 295(11) & 295(11) & 295(11) & 59(5) & 2.60(6)
\\  3DHC & 5 & [9.025, 0.219, 5.37, 5.57] & 143.5(4) & 867(16) & 248(19) & 51(3) & 250(11) & 250(11) & 2.60(8)
\\  3D & 6 & [6.128, 0, 6.128, 6.128] & 149.0(3) & 867(16) & 217(10) & 217(10) & 217(10) & 217(10) & 2.60(6)
\\ \hline
\end{tabular} 
\caption{Lattice parameters for each geometry. 
All lattice geometries of the measurements in Fig. 1 are shown with their individual lattice parameters: dimer, 1D chains, two-dimensional honeycomb planes (2DHC), two-leg ladder, square (2D), coupled honeycomb planes (3DHC) and simple cubic lattice (3D).
The different lattice geometries are listed in increasing order of the connectivity $Z$. 
The graph in Fig. \ref{fig_lattice_schematics} indicates the labeling of tunneling links used in the experiment. 
To set a constant value of the on-site interaction $U$ we vary the scattering length $a$. 
With increasing $Z$ the number of strong nearest-neighbor links is increasing accordingly (i. e. for the dimerized lattice the strong tunneling is $t_s/h = 635 \, \mathrm{Hz}$, while all other links have a much weaker tunneling rate of $130 \, \mathrm{Hz}$). 
Due to the exact choice of the individual tunneling links we ensure that the total bandwidth $W$ is constant. 
The respective errors are a result of the systematic uncertainty of our system parameters}
\label{table_lattice_configurations}
\end{center}
\end{table*}

\subsection{Fig. 2: Crossover between geometries} 

In the second set of measurements we explore the crossover between two geometries, either from a square lattice ($Z=4$) to 1D chains ($Z=2$), or we start with 1D chains which are then continuously deformed to a dimerized lattice ($Z=1$).  
In both cases we vary the connectivity by weakening half of the initially strong links to a value of $t$ while increasing the other links to $t_s$. 
At the same time the tunneling of the remaining weak links in the initial lattice geometry $t_w$ is modified accordingly to keep $t_s/t_w=5$ constant. 
To explain this in more detail we consider the case of the 1D to dimer crossover. 
Here $t_\mathrm{x}$ corresponds to the strong tunneling link $t_s$ and $t_\mathrm{x}'$ to $t$.
During the crossover $t_\mathrm{x}$ as one of the two initially strong links in the 1D chain ($t_\mathrm{x}$ and $t_\mathrm{x}'$) is increased, whereas $t_\mathrm{x}'$ is decreased to a lower value. 
Thereby the ratio between $t_\mathrm{x}$ and $t_\mathrm{x}'$ can be tuned from 1 to 5. 
Subsequently the tunneling links $t_\mathrm{y}$ and $t_\mathrm{z}$ (corresponding to $t_w$) are increased to keep their ratio with $t_\mathrm{x}$ constant.  
This procedure allows varying the strong-to-weak tunneling ratio $t_s/t$ from 1 to 5, thus allowing for a continuous scan between two geometries.  
As for the previous measurements we also keep the total bandwidth, the geometric mean trapping frequency and the on-site interaction constant. 

Each data point corresponds to a series of 25 measurements of the spin correlator after directly loading the fermions into the final lattice geometry. 
Due to the experimental set-up of the tunable optical lattice it is only possible to merge adjacent sites along the $x$-direction of the laboratory frame. 
In order to measure the spin correlations on neighboring sites along the perpendicular direction as shown in Fig. 2a, the atoms are loaded into a lattice geometry that is rotated by $90^{\circ}$, which is achieved by exchanging the values for the lattice depths $V_{\overline{X}}$ and $V_{Y}$.
By merging along the $x$-direction of the laboratory we then effectively merge along the perpendicular direction of the physical system. 
Due to the rotation of the lattice pattern the effective trapping frequencies along all directions of the harmonic confinement are different for the two lattice geometries. 
However, the local density approximation (LDA), which has been shown to be accurate in the temperature regime of our experiment \cite{Scarola2009}, entails that only the geometric mean of the trap frequency matters.   
This implies that the measured result does not depend on the rotation in the $xy$-plane at constant geometric mean trapping frequency.

To compare the spin correlations on weak and strong links in the dimerized lattice (see Fig. 2b) we need to be able to control which of the two sites we merge. 
This is achieved by controlling the phase $\varphi$ between the $X$ and $Y$ lattice beam.
During the time spent in the deep lattice (when $V_X=0)$, just before the merging ramp, the phase $\varphi$ is shifted by $\pi$. 
This ensures that we can merge two adjacent sites along the previously weak link. 
The phase shift is generated by an active feed-forward on the phase lock using the acousto-optical modulator of the lattice beam. 

\subsection{Fig. 3: Dynamics of spin correlations}

In order to observe dynamics in spin correlations we first ramp into the starting lattice geometry, which is either lattice of 1D chain in the $x$-direction ($Z=2$, with the same configuration as in Fig. 1) or a dimerized lattice ($Z=1$). 
In contrast to the previous measurements we then, in a second step, change the lattice geometry on a variable time scale $\tau$ before detecting the spin correlations. 

In the first measurement shown in Fig. 3a we change the 1D lattice with strong tunneling links $t_s$ initially along the $x$-direction to a geometry with 1D chains along the perpendicular direction. 
This is achieved by linearly ramping the lattice depths within the ramp time $\tau$ from $V_{\overline{X},X,Y,\widetilde{Z}}=[3.205, 0, 9.6, 9.6]\,E_R$ to $V_{\overline{X},X,Y,\widetilde{Z}}=[9.6, 0, 3.205, 9.6]\,E_R$ (at exactly half of the ramp time the potential is hence in a square lattice geometry with equal tunneling strengths in the $xy$-plane). 
The spin correlations are measured in the final geometry, where the strong tunneling links are now along the perpendicular direction. 
We do not allow for an additional wait time in the final lattice geometry, but directly ramp to the deep cubic lattice and follow the detection scheme described in the previous section.
To keep the time the atoms are exposed to the optical lattice constant, we add a wait time before the change in lattice geometry, such that the sum of the wait time and the lattice ramp time $\tau$ is $80\,\mathrm{ms}$. 
To measure correlations along new strong links we follow the same procedure as described in the section of the geometry scan.

In the second set of measurements shown in Fig. 3b we analyze the dynamics of spin correlations when linearly ramping from a dimerized lattice ($V_{\overline{X},X,Y,\widetilde{Z}}=[4.925, 0.073, 7.9, 8.19]\,E_R$) into a lattice with 1D chains along the $x$-direction ($V_{\overline{X},X,Y,\widetilde{Z}}=[3.205, 0, 9.6, 9.6]\,E_R$). 
We follow the same procedure as before with the only difference that the total time in the shallow lattice is now set to $100\,\mathrm{ms}$ to allow for increased ramp times $\tau$. 
To merge on different links (initially we have weak and strong tunneling links in the dimerized lattice) we follow the same procedure as described for the data in Fig. 2b.


\begin{thebibliography}{36}%
\makeatletter
\providecommand \@ifxundefined [1]{%
 \@ifx{#1\undefined}
}%
\providecommand \@ifnum [1]{%
 \ifnum #1\expandafter \@firstoftwo
 \else \expandafter \@secondoftwo
 \fi
}%
\providecommand \@ifx [1]{%
 \ifx #1\expandafter \@firstoftwo
 \else \expandafter \@secondoftwo
 \fi
}%
\providecommand \natexlab [1]{#1}%
\providecommand \enquote  [1]{``#1''}%
\providecommand \bibnamefont  [1]{#1}%
\providecommand \bibfnamefont [1]{#1}%
\providecommand \citenamefont [1]{#1}%
\providecommand \href@noop [0]{\@secondoftwo}%
\providecommand \href [0]{\begingroup \@sanitize@url \@href}%
\providecommand \@href[1]{\@@startlink{#1}\@@href}%
\providecommand \@@href[1]{\endgroup#1\@@endlink}%
\providecommand \@sanitize@url [0]{\catcode `\\12\catcode `\$12\catcode
  `\&12\catcode `\#12\catcode `\^12\catcode `\_12\catcode `\%12\relax}%
\providecommand \@@startlink[1]{}%
\providecommand \@@endlink[0]{}%
\providecommand \url  [0]{\begingroup\@sanitize@url \@url }%
\providecommand \@url [1]{\endgroup\@href {#1}{\urlprefix }}%
\providecommand \urlprefix  [0]{URL }%
\providecommand \Eprint [0]{\href }%
\providecommand \doibase [0]{http://dx.doi.org/}%
\providecommand \selectlanguage [0]{\@gobble}%
\providecommand \bibinfo  [0]{\@secondoftwo}%
\providecommand \bibfield  [0]{\@secondoftwo}%
\providecommand \translation [1]{[#1]}%
\providecommand \BibitemOpen [0]{}%
\providecommand \bibitemStop [0]{}%
\providecommand \bibitemNoStop [0]{.\EOS\space}%
\providecommand \EOS [0]{\spacefactor3000\relax}%
\providecommand \BibitemShut  [1]{\csname bibitem#1\endcsname}%
\let\auto@bib@innerbib\@empty
\bibitem [{\citenamefont {Auerbach}(1994)}]{auerbach1994}%
  \BibitemOpen
  \bibfield  {author} {\bibinfo {author} {\bibfnamefont {A.}~\bibnamefont
  {Auerbach}},\ }\href@noop {} {\emph {\bibinfo {title} {Interacting Electrons
  and Quantum Magnetism}}}\ (\bibinfo  {publisher} {Springer},\ \bibinfo {year}
  {1994})\BibitemShut {NoStop}%
\bibitem [{\citenamefont {Giamarchi}(2003)}]{giamarchi2003}%
  \BibitemOpen
  \bibfield  {author} {\bibinfo {author} {\bibfnamefont {T.}~\bibnamefont
  {Giamarchi}},\ }\href@noop {} {\emph {\bibinfo {title} {Quantum Physics in
  One Dimension}}}\ (\bibinfo  {publisher} {Clarendon Press},\ \bibinfo {year}
  {2003})\BibitemShut {NoStop}%
\bibitem [{\citenamefont {Balents}(2010)}]{balents2010}%
  \BibitemOpen
  \bibfield  {author} {\bibinfo {author} {\bibfnamefont {L.}~\bibnamefont
  {Balents}},\ }\href {\doibase 10.1038/nature08917} {\bibfield  {journal}
  {\bibinfo  {journal} {Nature}\ }\textbf {\bibinfo {volume} {464}},\ \bibinfo
  {pages} {199} (\bibinfo {year} {2010})}\BibitemShut {NoStop}%
\bibitem [{\citenamefont {Bloch}\ \emph {et~al.}(2008)\citenamefont {Bloch},
  \citenamefont {Dalibard},\ and\ \citenamefont {Zwerger}}]{Bloch2008d}%
  \BibitemOpen
  \bibfield  {author} {\bibinfo {author} {\bibfnamefont {I.}~\bibnamefont
  {Bloch}}, \bibinfo {author} {\bibfnamefont {J.}~\bibnamefont {Dalibard}}, \
  and\ \bibinfo {author} {\bibfnamefont {W.}~\bibnamefont {Zwerger}},\ }\href
  {\doibase 10.1103/RevModPhys.80.885} {\bibfield  {journal} {\bibinfo
  {journal} {Reviews of Modern Physics}\ }\textbf {\bibinfo {volume} {80}},\
  \bibinfo {pages} {885} (\bibinfo {year} {2008})}\BibitemShut {NoStop}%
\bibitem [{\citenamefont {Esslinger}(2010)}]{Esslinger2010c}%
  \BibitemOpen
  \bibfield  {author} {\bibinfo {author} {\bibfnamefont {T.}~\bibnamefont
  {Esslinger}},\ }\href {\doibase 10.1146/annurev-conmatphys-070909-104059}
  {\bibfield  {journal} {\bibinfo  {journal} {Annual Review of Condensed Matter
  Physics}\ }\textbf {\bibinfo {volume} {1}},\ \bibinfo {pages} {129} (\bibinfo
  {year} {2010})}\BibitemShut {NoStop}%
\bibitem [{\citenamefont {Sebby-Strabley}\ \emph {et~al.}(2006)\citenamefont
  {Sebby-Strabley}, \citenamefont {Anderlini}, \citenamefont {Jessen},\ and\
  \citenamefont {Porto}}]{Sebby-Strabley2006a}%
  \BibitemOpen
  \bibfield  {author} {\bibinfo {author} {\bibfnamefont {J.}~\bibnamefont
  {Sebby-Strabley}}, \bibinfo {author} {\bibfnamefont {M.}~\bibnamefont
  {Anderlini}}, \bibinfo {author} {\bibfnamefont {P.~S.}\ \bibnamefont
  {Jessen}}, \ and\ \bibinfo {author} {\bibfnamefont {J.~V.}\ \bibnamefont
  {Porto}},\ }\href {\doibase 10.1103/PhysRevA.73.033605} {\bibfield  {journal}
  {\bibinfo  {journal} {Physical Review A}\ }\textbf {\bibinfo {volume} {73}},\
  \bibinfo {pages} {033605} (\bibinfo {year} {2006})}\BibitemShut {NoStop}%
\bibitem [{\citenamefont {Soltan-Panahi}\ \emph {et~al.}(2011)\citenamefont
  {Soltan-Panahi}, \citenamefont {Struck}, \citenamefont {Hauke}, \citenamefont
  {Bick}, \citenamefont {Plenkers}, \citenamefont {Meineke}, \citenamefont
  {Becker}, \citenamefont {Windpassinger}, \citenamefont {Lewenstein},\ and\
  \citenamefont {Sengstock}}]{soltan-panahi2011}%
  \BibitemOpen
  \bibfield  {author} {\bibinfo {author} {\bibfnamefont {P.}~\bibnamefont
  {Soltan-Panahi}}, \bibinfo {author} {\bibfnamefont {J.}~\bibnamefont
  {Struck}}, \bibinfo {author} {\bibfnamefont {P.}~\bibnamefont {Hauke}},
  \bibinfo {author} {\bibfnamefont {A.}~\bibnamefont {Bick}}, \bibinfo {author}
  {\bibfnamefont {W.}~\bibnamefont {Plenkers}}, \bibinfo {author}
  {\bibfnamefont {G.}~\bibnamefont {Meineke}}, \bibinfo {author} {\bibfnamefont
  {C.}~\bibnamefont {Becker}}, \bibinfo {author} {\bibfnamefont
  {P.}~\bibnamefont {Windpassinger}}, \bibinfo {author} {\bibfnamefont
  {M.}~\bibnamefont {Lewenstein}}, \ and\ \bibinfo {author} {\bibfnamefont
  {K.}~\bibnamefont {Sengstock}},\ }\href {\doibase 10.1038/nphys1916}
  {\bibfield  {journal} {\bibinfo  {journal} {Nature Physics}\ }\textbf
  {\bibinfo {volume} {7}},\ \bibinfo {pages} {434} (\bibinfo {year}
  {2011})}\BibitemShut {NoStop}%
\bibitem [{\citenamefont {Tarruell}\ \emph {et~al.}(2012)\citenamefont
  {Tarruell}, \citenamefont {Greif}, \citenamefont {Uehlinger}, \citenamefont
  {Jotzu},\ and\ \citenamefont {Esslinger}}]{tarruell2012}%
  \BibitemOpen
  \bibfield  {author} {\bibinfo {author} {\bibfnamefont {L.}~\bibnamefont
  {Tarruell}}, \bibinfo {author} {\bibfnamefont {D.}~\bibnamefont {Greif}},
  \bibinfo {author} {\bibfnamefont {T.}~\bibnamefont {Uehlinger}}, \bibinfo
  {author} {\bibfnamefont {G.}~\bibnamefont {Jotzu}}, \ and\ \bibinfo {author}
  {\bibfnamefont {T.}~\bibnamefont {Esslinger}},\ }\href {\doibase 10.1038/nature10871} {\bibfield  {journal} {\bibinfo  {journal} {Nature}\
  }\textbf {\bibinfo {volume} {483}},\ \bibinfo {pages} {302} (\bibinfo {year}
  {2012})}\BibitemShut {NoStop}%
\bibitem [{\citenamefont {Jo}\ \emph {et~al.}(2012)\citenamefont {Jo},
  \citenamefont {Guzman}, \citenamefont {Thomas}, \citenamefont {Hosur},
  \citenamefont {Vishwanath},\ and\ \citenamefont {Stamper-Kurn}}]{jo2012}%
  \BibitemOpen
  \bibfield  {author} {\bibinfo {author} {\bibfnamefont {G.-B.}\ \bibnamefont
  {Jo}}, \bibinfo {author} {\bibfnamefont {J.}~\bibnamefont {Guzman}}, \bibinfo
  {author} {\bibfnamefont {C.~K.}\ \bibnamefont {Thomas}}, \bibinfo {author}
  {\bibfnamefont {P.}~\bibnamefont {Hosur}}, \bibinfo {author} {\bibfnamefont
  {A.}~\bibnamefont {Vishwanath}}, \ and\ \bibinfo {author} {\bibfnamefont
  {D.~M.}\ \bibnamefont {Stamper-Kurn}},\ }\href {\doibase 10.1103/PhysRevLett.108.045305} {\bibfield  {journal} {\bibinfo  {journal}
  {Physical Review Letters}\ }\textbf {\bibinfo {volume} {108}},\ \bibinfo
  {pages} {045305} (\bibinfo {year} {2012})}\BibitemShut {NoStop}%
\bibitem [{\citenamefont {Taie}\ \emph {et~al.}(2015)\citenamefont {Taie},
  \citenamefont {Ozawa}, \citenamefont {Ichinose}, \citenamefont {Nishio},
  \citenamefont {Nakajima},\ and\ \citenamefont {Takahashi}}]{Taie2015}%
  \BibitemOpen
  \bibfield  {author} {\bibinfo {author} {\bibfnamefont {S.}~\bibnamefont
  {Taie}}, \bibinfo {author} {\bibfnamefont {H.}~\bibnamefont {Ozawa}},
  \bibinfo {author} {\bibfnamefont {T.}~\bibnamefont {Ichinose}}, \bibinfo
  {author} {\bibfnamefont {T.}~\bibnamefont {Nishio}}, \bibinfo {author}
  {\bibfnamefont {S.}~\bibnamefont {Nakajima}}, \ and\ \bibinfo {author}
  {\bibfnamefont {Y.}~\bibnamefont {Takahashi}},\ }\href
  {http://arxiv.org/abs/1506.00587} {\bibfield  {journal} {\bibinfo  {journal}
  {arXiv:1506.00587}\ } (\bibinfo {year} {2015})}\BibitemShut {NoStop}%
\bibitem [{\citenamefont {Yan}\ \emph {et~al.}(2013)\citenamefont {Yan},
  \citenamefont {Moses}, \citenamefont {Gadway}, \citenamefont {Covey},
  \citenamefont {Hazzard}, \citenamefont {Rey}, \citenamefont {Jin},\ and\
  \citenamefont {Ye}}]{Yan2013}%
  \BibitemOpen
  \bibfield  {author} {\bibinfo {author} {\bibfnamefont {B.}~\bibnamefont
  {Yan}}, \bibinfo {author} {\bibfnamefont {S.~A.}\ \bibnamefont {Moses}},
  \bibinfo {author} {\bibfnamefont {B.}~\bibnamefont {Gadway}}, \bibinfo
  {author} {\bibfnamefont {J.~P.}\ \bibnamefont {Covey}}, \bibinfo {author}
  {\bibfnamefont {K.~R.~A.}\ \bibnamefont {Hazzard}}, \bibinfo {author}
  {\bibfnamefont {A.~M.}\ \bibnamefont {Rey}}, \bibinfo {author} {\bibfnamefont
  {D.~S.}\ \bibnamefont {Jin}}, \ and\ \bibinfo {author} {\bibfnamefont
  {J.}~\bibnamefont {Ye}},\ }\href {\doibase 10.1038/nature12483} {\bibfield
  {journal} {\bibinfo  {journal} {Nature}\ }\textbf {\bibinfo {volume} {501}},\
  \bibinfo {pages} {521} (\bibinfo {year} {2013})}\BibitemShut {NoStop}%
\bibitem [{\citenamefont {Greif}\ \emph {et~al.}(2013)\citenamefont {Greif},
  \citenamefont {Uehlinger}, \citenamefont {Jotzu}, \citenamefont {Tarruell},\
  and\ \citenamefont {Esslinger}}]{greif2013}%
  \BibitemOpen
  \bibfield  {author} {\bibinfo {author} {\bibfnamefont {D.}~\bibnamefont
  {Greif}}, \bibinfo {author} {\bibfnamefont {T.}~\bibnamefont {Uehlinger}},
  \bibinfo {author} {\bibfnamefont {G.}~\bibnamefont {Jotzu}}, \bibinfo
  {author} {\bibfnamefont {L.}~\bibnamefont {Tarruell}}, \ and\ \bibinfo
  {author} {\bibfnamefont {T.}~\bibnamefont {Esslinger}},\ }\href {\doibase 10.1126/science.1236362} {\bibfield  {journal} {\bibinfo  {journal}
  {Science}\ }\textbf {\bibinfo {volume} {340}},\ \bibinfo {pages} {1307}
  (\bibinfo {year} {2013})}\BibitemShut {NoStop}%
\bibitem [{\citenamefont {Hart}\ \emph {et~al.}(2015)\citenamefont {Hart},
  \citenamefont {Duarte}, \citenamefont {Yang}, \citenamefont {Liu},
  \citenamefont {Paiva}, \citenamefont {Khatami}, \citenamefont {Scalettar},
  \citenamefont {Trivedi}, \citenamefont {Huse},\ and\ \citenamefont
  {Hulet}}]{Hart2015}%
  \BibitemOpen
  \bibfield  {author} {\bibinfo {author} {\bibfnamefont {R.~A.}\ \bibnamefont
  {Hart}}, \bibinfo {author} {\bibfnamefont {P.~M.}\ \bibnamefont {Duarte}},
  \bibinfo {author} {\bibfnamefont {T.-L.}\ \bibnamefont {Yang}}, \bibinfo
  {author} {\bibfnamefont {X.}~\bibnamefont {Liu}}, \bibinfo {author}
  {\bibfnamefont {T.}~\bibnamefont {Paiva}}, \bibinfo {author} {\bibfnamefont
  {E.}~\bibnamefont {Khatami}}, \bibinfo {author} {\bibfnamefont {R.~T.}\
  \bibnamefont {Scalettar}}, \bibinfo {author} {\bibfnamefont {N.}~\bibnamefont
  {Trivedi}}, \bibinfo {author} {\bibfnamefont {D.~A.}\ \bibnamefont {Huse}}, \
  and\ \bibinfo {author} {\bibfnamefont {R.~G.}\ \bibnamefont {Hulet}},\ }\href
  {\doibase 10.1038/nature14223} {\bibfield  {journal} {\bibinfo  {journal}
  {Nature}\ }\textbf {\bibinfo {volume} {519}},\ \bibinfo {pages} {211}
  (\bibinfo {year} {2015})}\BibitemShut {NoStop}%
\bibitem [{\citenamefont {Sciolla}\ \emph {et~al.}(2013)\citenamefont
  {Sciolla}, \citenamefont {Tokuno}, \citenamefont {Uchino}, \citenamefont
  {Barmettler}, \citenamefont {Giamarchi},\ and\ \citenamefont
  {Kollath}}]{Sciolla2013}%
  \BibitemOpen
  \bibfield  {author} {\bibinfo {author} {\bibfnamefont {B.}~\bibnamefont
  {Sciolla}}, \bibinfo {author} {\bibfnamefont {A.}~\bibnamefont {Tokuno}},
  \bibinfo {author} {\bibfnamefont {S.}~\bibnamefont {Uchino}}, \bibinfo
  {author} {\bibfnamefont {P.}~\bibnamefont {Barmettler}}, \bibinfo {author}
  {\bibfnamefont {T.}~\bibnamefont {Giamarchi}}, \ and\ \bibinfo {author}
  {\bibfnamefont {C.}~\bibnamefont {Kollath}},\ }\href {\doibase 10.1103/PhysRevA.88.063629} {\bibfield  {journal} {\bibinfo  {journal}
  {Physical Review A}\ }\textbf {\bibinfo {volume} {88}},\ \bibinfo {pages}
  {063629} (\bibinfo {year} {2013})}\BibitemShut {NoStop}%
\bibitem [{\citenamefont {{Imri\v ska}}\ \emph {et~al.}(2014)\citenamefont
  {{Imri\v ska}}, \citenamefont {Iazzi}, \citenamefont {Wang}, \citenamefont
  {Gull}, \citenamefont {Greif}, \citenamefont {Uehlinger}, \citenamefont
  {Jotzu}, \citenamefont {Tarruell}, \citenamefont {Esslinger},\ and\
  \citenamefont {Troyer}}]{imriska2014}%
  \BibitemOpen
  \bibfield  {author} {\bibinfo {author} {\bibfnamefont {J.}~\bibnamefont
  {{Imri\v ska}}}, \bibinfo {author} {\bibfnamefont {M.}~\bibnamefont {Iazzi}},
  \bibinfo {author} {\bibfnamefont {L.}~\bibnamefont {Wang}}, \bibinfo {author}
  {\bibfnamefont {E.}~\bibnamefont {Gull}}, \bibinfo {author} {\bibfnamefont
  {D.}~\bibnamefont {Greif}}, \bibinfo {author} {\bibfnamefont
  {T.}~\bibnamefont {Uehlinger}}, \bibinfo {author} {\bibfnamefont
  {G.}~\bibnamefont {Jotzu}}, \bibinfo {author} {\bibfnamefont
  {L.}~\bibnamefont {Tarruell}}, \bibinfo {author} {\bibfnamefont
  {T.}~\bibnamefont {Esslinger}}, \ and\ \bibinfo {author} {\bibfnamefont
  {M.}~\bibnamefont {Troyer}},\ }\href {\doibase 10.1103/PhysRevLett.112.115301} {\bibfield  {journal} {\bibinfo  {journal}
  {Physical Review Letters}\ }\textbf {\bibinfo {volume} {112}},\ \bibinfo
  {pages} {115301} (\bibinfo {year} {2014})}\BibitemShut {NoStop}%
\bibitem [{\citenamefont {Golubeva}\ \emph {et~al.}(2015)\citenamefont {Golubeva},
  \citenamefont {Sotnikov},\ and\ \citenamefont {Hofstetter}}]{golubeva2015}%
  \BibitemOpen
  \bibfield  {author} {\bibinfo {author} {\bibfnamefont {A.}~\bibnamefont
  {Golubeva}}, \bibinfo {author} {\bibfnamefont {A.}\ \bibnamefont {Sotnikov}},\ 
	and\ \bibinfo {author} {\bibfnamefont {W.}~\bibnamefont {Hofstetter}},\ }\href 
	{http://arxiv.org/abs/1505.02733} {\bibfield  {journal} {\bibinfo
  {journal} {arXiv:1505.02733}\ } (\bibinfo {year} {2015})}\BibitemShut
  {NoStop}%
\bibitem [{\citenamefont {Simon}\ \emph {et~al.}(2011)\citenamefont {Simon},
  \citenamefont {Bakr}, \citenamefont {Ma}, \citenamefont {Tai}, \citenamefont
  {Preiss},\ and\ \citenamefont {Greiner}}]{simon2011}%
  \BibitemOpen
  \bibfield  {author} {\bibinfo {author} {\bibfnamefont {J.}~\bibnamefont
  {Simon}}, \bibinfo {author} {\bibfnamefont {W.~S.}\ \bibnamefont {Bakr}},
  \bibinfo {author} {\bibfnamefont {R.}~\bibnamefont {Ma}}, \bibinfo {author}
  {\bibfnamefont {M.~E.}\ \bibnamefont {Tai}}, \bibinfo {author} {\bibfnamefont
  {P.~M.}\ \bibnamefont {Preiss}}, \ and\ \bibinfo {author} {\bibfnamefont
  {M.}~\bibnamefont {Greiner}},\ }\href {\doibase 10.1038/nature09994}
  {\bibfield  {journal} {\bibinfo  {journal} {Nature}\ }\textbf {\bibinfo
  {volume} {472}},\ \bibinfo {pages} {307} (\bibinfo {year}
  {2011})}\BibitemShut {NoStop}%
\bibitem [{\citenamefont {Struck}\ \emph {et~al.}(2011)\citenamefont {Struck},
  \citenamefont {\"Olschl\"ager}, \citenamefont {Targat}, \citenamefont
  {Soltan-Panahi}, \citenamefont {Eckardt}, \citenamefont {Lewenstein},
  \citenamefont {Windpassinger},\ and\ \citenamefont {Sengstock}}]{struck2011}%
  \BibitemOpen
  \bibfield  {author} {\bibinfo {author} {\bibfnamefont {J.}~\bibnamefont
  {Struck}}, \bibinfo {author} {\bibfnamefont {C.}~\bibnamefont
  {\"Olschl\"ager}}, \bibinfo {author} {\bibfnamefont {R.~L.}\ \bibnamefont
  {Targat}}, \bibinfo {author} {\bibfnamefont {P.}~\bibnamefont
  {Soltan-Panahi}}, \bibinfo {author} {\bibfnamefont {A.}~\bibnamefont
  {Eckardt}}, \bibinfo {author} {\bibfnamefont {M.}~\bibnamefont {Lewenstein}},
  \bibinfo {author} {\bibfnamefont {P.}~\bibnamefont {Windpassinger}}, \ and\
  \bibinfo {author} {\bibfnamefont {K.}~\bibnamefont {Sengstock}},\ }\href
  {\doibase 10.1126/science.1207239} {\bibfield  {journal} {\bibinfo  {journal}
  {Science}\ }\textbf {\bibinfo {volume} {333}},\ \bibinfo {pages} {996}
  (\bibinfo {year} {2011})}\BibitemShut {NoStop}%
\bibitem [{\citenamefont {Struck}\ \emph {et~al.}(2013)\citenamefont {Struck},
  \citenamefont {Weinberg}, \citenamefont {\"Olschl\"ager}, \citenamefont
  {Windpassinger}, \citenamefont {Simonet}, \citenamefont {Sengstock},
  \citenamefont {H\"oppner}, \citenamefont {Hauke}, \citenamefont {Eckardt},
  \citenamefont {Lewenstein},\ and\ \citenamefont {Mathey}}]{struck2013}%
  \BibitemOpen
  \bibfield  {author} {\bibinfo {author} {\bibfnamefont {J.}~\bibnamefont
  {Struck}}, \bibinfo {author} {\bibfnamefont {M.}~\bibnamefont {Weinberg}},
  \bibinfo {author} {\bibfnamefont {C.}~\bibnamefont {\"Olschl\"ager}},
  \bibinfo {author} {\bibfnamefont {P.}~\bibnamefont {Windpassinger}}, \bibinfo
  {author} {\bibfnamefont {J.}~\bibnamefont {Simonet}}, \bibinfo {author}
  {\bibfnamefont {K.}~\bibnamefont {Sengstock}}, \bibinfo {author}
  {\bibfnamefont {R.}~\bibnamefont {H\"oppner}}, \bibinfo {author}
  {\bibfnamefont {P.}~\bibnamefont {Hauke}}, \bibinfo {author} {\bibfnamefont
  {A.}~\bibnamefont {Eckardt}}, \bibinfo {author} {\bibfnamefont
  {M.}~\bibnamefont {Lewenstein}}, \ and\ \bibinfo {author} {\bibfnamefont
  {L.}~\bibnamefont {Mathey}},\ }\href {\doibase 10.1038/nphys2750} {\bibfield
  {journal} {\bibinfo  {journal} {Nature Physics}\ }\textbf {\bibinfo {volume}
  {9}},\ \bibinfo {pages} {738} (\bibinfo {year} {2013})}\BibitemShut {NoStop}%
\bibitem [{\citenamefont {Fukuhara}\ \emph
  {et~al.}(2013{\natexlab{a}})\citenamefont {Fukuhara}, \citenamefont
  {Schau\ss}, \citenamefont {Endres}, \citenamefont {Hild}, \citenamefont
  {Cheneau}, \citenamefont {Bloch},\ and\ \citenamefont
  {Gross}}]{fukuhara2013}%
  \BibitemOpen
  \bibfield  {author} {\bibinfo {author} {\bibfnamefont {T.}~\bibnamefont
  {Fukuhara}}, \bibinfo {author} {\bibfnamefont {P.}~\bibnamefont {Schau\ss}},
  \bibinfo {author} {\bibfnamefont {M.}~\bibnamefont {Endres}}, \bibinfo
  {author} {\bibfnamefont {S.}~\bibnamefont {Hild}}, \bibinfo {author}
  {\bibfnamefont {M.}~\bibnamefont {Cheneau}}, \bibinfo {author} {\bibfnamefont
  {I.}~\bibnamefont {Bloch}}, \ and\ \bibinfo {author} {\bibfnamefont
  {C.}~\bibnamefont {Gross}},\ }\href {\doibase 10.1038/nature12541} {\bibfield
   {journal} {\bibinfo  {journal} {Nature}\ }\textbf {\bibinfo {volume}
  {502}},\ \bibinfo {pages} {76} (\bibinfo {year}
  {2013}{\natexlab{a}})}\BibitemShut {NoStop}%
\bibitem [{\citenamefont {Fukuhara}\ \emph
  {et~al.}(2013{\natexlab{b}})\citenamefont {Fukuhara}, \citenamefont
  {Kantian}, \citenamefont {Endres}, \citenamefont {Cheneau}, \citenamefont
  {Schau\ss}, \citenamefont {Hild}, \citenamefont {Bellem}, \citenamefont
  {Schollw\"ock}, \citenamefont {Giamarchi}, \citenamefont {Gross},
  \citenamefont {Bloch},\ and\ \citenamefont {Kuhr}}]{fukuhara2013b}%
  \BibitemOpen
  \bibfield  {author} {\bibinfo {author} {\bibfnamefont {T.}~\bibnamefont
  {Fukuhara}}, \bibinfo {author} {\bibfnamefont {A.}~\bibnamefont {Kantian}},
  \bibinfo {author} {\bibfnamefont {M.}~\bibnamefont {Endres}}, \bibinfo
  {author} {\bibfnamefont {M.}~\bibnamefont {Cheneau}}, \bibinfo {author}
  {\bibfnamefont {P.}~\bibnamefont {Schau\ss}}, \bibinfo {author}
  {\bibfnamefont {S.}~\bibnamefont {Hild}}, \bibinfo {author} {\bibfnamefont
  {D.}~\bibnamefont {Bellem}}, \bibinfo {author} {\bibfnamefont
  {U.}~\bibnamefont {Schollw\"ock}}, \bibinfo {author} {\bibfnamefont
  {T.}~\bibnamefont {Giamarchi}}, \bibinfo {author} {\bibfnamefont
  {C.}~\bibnamefont {Gross}}, \bibinfo {author} {\bibfnamefont
  {I.}~\bibnamefont {Bloch}}, \ and\ \bibinfo {author} {\bibfnamefont
  {S.}~\bibnamefont {Kuhr}},\ }\href {\doibase 10.1038/nphys2561} {\bibfield
  {journal} {\bibinfo  {journal} {Nature Physics}\ }\textbf {\bibinfo {volume}
  {9}},\ \bibinfo {pages} {235} (\bibinfo {year}
  {2013}{\natexlab{b}})}\BibitemShut {NoStop}%
\bibitem [{\citenamefont {Hild}\ \emph {et~al.}(2014)\citenamefont {Hild},
  \citenamefont {Fukuhara}, \citenamefont {Schau\ss}, \citenamefont {Zeiher},
  \citenamefont {Knap}, \citenamefont {Demler}, \citenamefont {Bloch},\ and\
  \citenamefont {Gross}}]{Hild2014}%
  \BibitemOpen
  \bibfield  {author} {\bibinfo {author} {\bibfnamefont {S.}~\bibnamefont
  {Hild}}, \bibinfo {author} {\bibfnamefont {T.}~\bibnamefont {Fukuhara}},
  \bibinfo {author} {\bibfnamefont {P.}~\bibnamefont {Schau\ss}}, \bibinfo
  {author} {\bibfnamefont {J.}~\bibnamefont {Zeiher}}, \bibinfo {author}
  {\bibfnamefont {M.}~\bibnamefont {Knap}}, \bibinfo {author} {\bibfnamefont
  {E.}~\bibnamefont {Demler}}, \bibinfo {author} {\bibfnamefont
  {I.}~\bibnamefont {Bloch}}, \ and\ \bibinfo {author} {\bibfnamefont
  {C.}~\bibnamefont {Gross}},\ }\href {\doibase 10.1103/PhysRevLett.113.147205}
  {\bibfield  {journal} {\bibinfo  {journal} {Physical Review Letters}\
  }\textbf {\bibinfo {volume} {113}},\ \bibinfo {pages} {147205} (\bibinfo
  {year} {2014})}\BibitemShut {NoStop}%
\bibitem [{\citenamefont {Brown}\ \emph {et~al.}(2015)\citenamefont {Brown},
  \citenamefont {Wyllie}, \citenamefont {Koller}, \citenamefont {Goldschmidt},
  \citenamefont {Foss-Feig},\ and\ \citenamefont {Porto}}]{Brown2015}%
  \BibitemOpen
  \bibfield  {author} {\bibinfo {author} {\bibfnamefont {R.~C.}\ \bibnamefont
  {Brown}}, \bibinfo {author} {\bibfnamefont {R.}~\bibnamefont {Wyllie}},
  \bibinfo {author} {\bibfnamefont {S.~B.}\ \bibnamefont {Koller}}, \bibinfo
  {author} {\bibfnamefont {E.~A.}\ \bibnamefont {Goldschmidt}}, \bibinfo
  {author} {\bibfnamefont {M.}~\bibnamefont {Foss-Feig}}, \ and\ \bibinfo
  {author} {\bibfnamefont {J.~V.}\ \bibnamefont {Porto}},\ }\href {\doibase 10.1126/science.aaa1385} {\bibfield  {journal} {\bibinfo  {journal} {Science
  (New York, N.Y.)}\ }\textbf {\bibinfo {volume} {348}},\ \bibinfo {pages}
  {540} (\bibinfo {year} {2015})}\BibitemShut {NoStop}%
\bibitem [{\citenamefont {See-Supplemental-Information}()}]{si}%
  \BibitemOpen
  \bibinfo {note} {See Supplemental Material for details}
	\BibitemShut {NoStop}%
\bibitem [{Note1()}]{Note1}%
  \BibitemOpen
  \bibinfo {note} {All Hubbard parameters are calculated from the lattice
  potential using Wannier functions, which are obtained using band-projected
  position operators \cite {uehlinger2013}.}\BibitemShut {Stop}%
\bibitem [{\citenamefont {Uehlinger}\ \emph {et~al.}(2013)\citenamefont
  {Uehlinger}, \citenamefont {Jotzu}, \citenamefont {Messer}, \citenamefont
  {Greif}, \citenamefont {Hofstetter}, \citenamefont {Bissbort},\ and\
  \citenamefont {Esslinger}}]{uehlinger2013}%
  \BibitemOpen
  \bibfield  {author} {\bibinfo {author} {\bibfnamefont {T.}~\bibnamefont
  {Uehlinger}}, \bibinfo {author} {\bibfnamefont {G.}~\bibnamefont {Jotzu}},
  \bibinfo {author} {\bibfnamefont {M.}~\bibnamefont {Messer}}, \bibinfo
  {author} {\bibfnamefont {D.}~\bibnamefont {Greif}}, \bibinfo {author}
  {\bibfnamefont {W.}~\bibnamefont {Hofstetter}}, \bibinfo {author}
  {\bibfnamefont {U.}~\bibnamefont {Bissbort}}, \ and\ \bibinfo {author}
  {\bibfnamefont {T.}~\bibnamefont {Esslinger}},\ }\href {\doibase 10.1103/PhysRevLett.111.185307} {\bibfield  {journal} {\bibinfo  {journal}
  {Physical Review Letters}\ }\textbf {\bibinfo {volume} {111}},\ \bibinfo
  {pages} {185307} (\bibinfo {year} {2013})}\BibitemShut {NoStop}%
\bibitem [{\citenamefont {Gorelik}\ \emph {et~al.}(2012)\citenamefont
  {Gorelik}, \citenamefont {Rost}, \citenamefont {Paiva}, \citenamefont
  {Scalettar}, \citenamefont {Kl\"{u}mper},\ and\ \citenamefont
  {Bl\"{u}mer}}]{Gorelik2012}%
  \BibitemOpen
  \bibfield  {author} {\bibinfo {author} {\bibfnamefont {E.~V.}\ \bibnamefont
  {Gorelik}}, \bibinfo {author} {\bibfnamefont {D.}~\bibnamefont {Rost}},
  \bibinfo {author} {\bibfnamefont {T.}~\bibnamefont {Paiva}}, \bibinfo
  {author} {\bibfnamefont {R.}~\bibnamefont {Scalettar}}, \bibinfo {author}
  {\bibfnamefont {A.}~\bibnamefont {Kl\"{u}mper}}, \ and\ \bibinfo {author}
  {\bibfnamefont {N.}~\bibnamefont {Bl\"{u}mer}},\ }\href {\doibase 10.1103/PhysRevA.85.061602} {\bibfield  {journal} {\bibinfo  {journal}
  {Physical Review A}\ }\textbf {\bibinfo {volume} {85}},\ \bibinfo {pages}
  {061602} (\bibinfo {year} {2012})}\BibitemShut {NoStop}%
\bibitem [{\citenamefont {Jotzu}\ \emph {et~al.}(2015)\citenamefont {Jotzu},
  \citenamefont {Messer}, \citenamefont {G\"{o}rg}, \citenamefont {Greif},
  \citenamefont {Desbuquois},\ and\ \citenamefont {Esslinger}}]{Jotzu2015}%
  \BibitemOpen
  \bibfield  {author} {\bibinfo {author} {\bibfnamefont {G.}~\bibnamefont
  {Jotzu}}, \bibinfo {author} {\bibfnamefont {M.}~\bibnamefont {Messer}},
  \bibinfo {author} {\bibfnamefont {F.}~\bibnamefont {G\"{o}rg}}, \bibinfo
  {author} {\bibfnamefont {D.}~\bibnamefont {Greif}}, \bibinfo {author}
  {\bibfnamefont {R.}~\bibnamefont {Desbuquois}}, \ and\ \bibinfo {author}
  {\bibfnamefont {T.}~\bibnamefont {Esslinger}},\ }\href
  {http://link.aps.org/doi/10.1103/PhysRevLett.115.073002} {\bibfield
  {journal} {\bibinfo  {journal} {Physical Review Letters}\ }\textbf {\bibinfo
  {volume} {115}},\ \bibinfo {pages} {073002} (\bibinfo {year}
  {2015})}\BibitemShut {NoStop}%
\bibitem [{\citenamefont {Freericks}\ and\ \citenamefont
  {Zlati\'{c}}(2003)}]{Freericks2003}%
  \BibitemOpen
  \bibfield  {author} {\bibinfo {author} {\bibfnamefont {J.~K.}\ \bibnamefont
  {Freericks}}\ and\ \bibinfo {author} {\bibfnamefont {V.}~\bibnamefont
  {Zlati\'{c}}},\ }\href {\doibase 10.1103/RevModPhys.75.1333} {\bibfield
  {journal} {\bibinfo  {journal} {Reviews of Modern Physics}\ }\textbf
  {\bibinfo {volume} {75}},\ \bibinfo {pages} {1333} (\bibinfo {year}
  {2003})}\BibitemShut {NoStop}%
\bibitem [{\citenamefont {Polkovnikov}\ \emph {et~al.}(2011)\citenamefont
  {Polkovnikov}, \citenamefont {Sengupta}, \citenamefont {Silva},\ and\
  \citenamefont {Vengalattore}}]{Polkovnikov2011a}%
  \BibitemOpen
  \bibfield  {author} {\bibinfo {author} {\bibfnamefont {A.}~\bibnamefont
  {Polkovnikov}}, \bibinfo {author} {\bibfnamefont {K.}~\bibnamefont
  {Sengupta}}, \bibinfo {author} {\bibfnamefont {A.}~\bibnamefont {Silva}}, \
  and\ \bibinfo {author} {\bibfnamefont {M.}~\bibnamefont {Vengalattore}},\
  }\href {\doibase 10.1103/RevModPhys.83.863} {\bibfield  {journal} {\bibinfo
  {journal} {Reviews of Modern Physics}\ }\textbf {\bibinfo {volume} {83}},\
  \bibinfo {pages} {863} (\bibinfo {year} {2011})}\BibitemShut {NoStop}%
\bibitem [{\citenamefont {Eisert}\ \emph {et~al.}(2015)\citenamefont {Eisert},
  \citenamefont {Friesdorf},\ and\ \citenamefont {Gogolin}}]{Eisert2015}%
  \BibitemOpen
  \bibfield  {author} {\bibinfo {author} {\bibfnamefont {J.}~\bibnamefont
  {Eisert}}, \bibinfo {author} {\bibfnamefont {M.}~\bibnamefont {Friesdorf}}, \
  and\ \bibinfo {author} {\bibfnamefont {C.}~\bibnamefont {Gogolin}},\ }\href
  {\doibase 10.1038/nphys3215} {\bibfield  {journal} {\bibinfo  {journal}
  {Nature Physics}\ }\textbf {\bibinfo {volume} {11}},\ \bibinfo {pages} {124}
  (\bibinfo {year} {2015})}\BibitemShut {NoStop}%
\bibitem [{\citenamefont {Ho}\ and\ \citenamefont {Zhou}(2009)}]{Ho2009a}%
  \BibitemOpen
  \bibfield  {author} {\bibinfo {author} {\bibfnamefont {T.-L.}\ \bibnamefont
  {Ho}}\ and\ \bibinfo {author} {\bibfnamefont {Q.}~\bibnamefont {Zhou}},\
  }\href {http://arxiv.org/abs/0911.5506} {\bibfield  {journal} {\bibinfo
  {journal} {arXiv:0911.5506}\ } (\bibinfo {year} {2009})}\BibitemShut
  {NoStop}%
\bibitem [{\citenamefont {Bernier}\ \emph {et~al.}(2009)\citenamefont
  {Bernier}, \citenamefont {Kollath}, \citenamefont {Georges}, \citenamefont
  {De~Leo}, \citenamefont {Gerbier}, \citenamefont {Salomon},\ and\
  \citenamefont {K\"ohl}}]{bernier2009}%
  \BibitemOpen
  \bibfield  {author} {\bibinfo {author} {\bibfnamefont {J.-S.}\ \bibnamefont
  {Bernier}}, \bibinfo {author} {\bibfnamefont {C.}~\bibnamefont {Kollath}},
  \bibinfo {author} {\bibfnamefont {A.}~\bibnamefont {Georges}}, \bibinfo
  {author} {\bibfnamefont {L.}~\bibnamefont {De~Leo}}, \bibinfo {author}
  {\bibfnamefont {F.}~\bibnamefont {Gerbier}}, \bibinfo {author} {\bibfnamefont
  {C.}~\bibnamefont {Salomon}}, \ and\ \bibinfo {author} {\bibfnamefont
  {M.}~\bibnamefont {K\"ohl}},\ }\href {\doibase 10.1103/PhysRevA.79.061601}
  {\bibfield  {journal} {\bibinfo  {journal} {Physical Review A}\ }\textbf
  {\bibinfo {volume} {79}},\ \bibinfo {pages} {061601} (\bibinfo {year}
  {2009})}\BibitemShut {NoStop}%
\bibitem [{\citenamefont {Lubasch}\ \emph {et~al.}(2011)\citenamefont
  {Lubasch}, \citenamefont {Murg}, \citenamefont {Schneider}, \citenamefont
  {Cirac},\ and\ \citenamefont {Ba\~nuls}}]{lubasch2011}%
  \BibitemOpen
  \bibfield  {author} {\bibinfo {author} {\bibfnamefont {M.}~\bibnamefont
  {Lubasch}}, \bibinfo {author} {\bibfnamefont {V.}~\bibnamefont {Murg}},
  \bibinfo {author} {\bibfnamefont {U.}~\bibnamefont {Schneider}}, \bibinfo
  {author} {\bibfnamefont {J.~I.}\ \bibnamefont {Cirac}}, \ and\ \bibinfo
  {author} {\bibfnamefont {M.-C.}\ \bibnamefont {Ba\~nuls}},\ }\href {\doibase 10.1103/PhysRevLett.107.165301} {\bibfield  {journal} {\bibinfo  {journal}
  {Physical Review Letters}\ }\textbf {\bibinfo {volume} {107}},\ \bibinfo
  {pages} {165301} (\bibinfo {year} {2011})}\BibitemShut {NoStop}%
\end{thebibliography}

\begin{thebibliography}{45}%
\makeatletter
\providecommand \@ifxundefined [1]{%
 \@ifx{#1\undefined}
}%
\providecommand \@ifnum [1]{%
 \ifnum #1\expandafter \@firstoftwo
 \else \expandafter \@secondoftwo
 \fi
}%
\providecommand \@ifx [1]{%
 \ifx #1\expandafter \@firstoftwo
 \else \expandafter \@secondoftwo
 \fi
}%
\providecommand \natexlab [1]{#1}%
\providecommand \enquote  [1]{``#1''}%
\providecommand \bibnamefont  [1]{#1}%
\providecommand \bibfnamefont [1]{#1}%
\providecommand \citenamefont [1]{#1}%
\providecommand \href@noop [0]{\@secondoftwo}%
\providecommand \href [0]{\begingroup \@sanitize@url \@href}%
\providecommand \@href[1]{\@@startlink{#1}\@@href}%
\providecommand \@@href[1]{\endgroup#1\@@endlink}%
\providecommand \@sanitize@url [0]{\catcode `\\12\catcode `\$12\catcode
  `\&12\catcode `\#12\catcode `\^12\catcode `\_12\catcode `\%12\relax}%
\providecommand \@@startlink[1]{}%
\providecommand \@@endlink[0]{}%
\providecommand \url  [0]{\begingroup\@sanitize@url \@url }%
\providecommand \@url [1]{\endgroup\@href {#1}{\urlprefix }}%
\providecommand \urlprefix  [0]{URL }%
\providecommand \Eprint [0]{\href }%
\providecommand \doibase [0]{http://dx.doi.org/}%
\providecommand \selectlanguage [0]{\@gobble}%
\providecommand \bibinfo  [0]{\@secondoftwo}%
\providecommand \bibfield  [0]{\@secondoftwo}%
\providecommand \translation [1]{[#1]}%
\providecommand \BibitemOpen [0]{}%
\providecommand \bibitemStop [0]{}%
\providecommand \bibitemNoStop [0]{.\EOS\space}%
\providecommand \EOS [0]{\spacefactor3000\relax}%
\providecommand \BibitemShut  [1]{\csname bibitem#1\endcsname}%
\let\auto@bib@innerbib\@empty
\bibitem [{\citenamefont {Tarruell}\ \emph {et~al.}(2012)\citenamefont
  {Tarruell}, \citenamefont {Greif}, \citenamefont {Uehlinger}, \citenamefont
  {Jotzu},\ and\ \citenamefont {Esslinger}}]{Tarruell2012}%
  \BibitemOpen
  \bibfield  {author} {\bibinfo {author} {\bibfnamefont {L.}~\bibnamefont
  {Tarruell}}, \bibinfo {author} {\bibfnamefont {D.}~\bibnamefont {Greif}},
  \bibinfo {author} {\bibfnamefont {T.}~\bibnamefont {Uehlinger}}, \bibinfo
  {author} {\bibfnamefont {G.}~\bibnamefont {Jotzu}}, \ and\ \bibinfo {author}
  {\bibfnamefont {T.}~\bibnamefont {Esslinger}},\ }\href {\doibase 10.1038/nature10871} {\bibfield  {journal} {\bibinfo  {journal} {Nature}\
  }\textbf {\bibinfo {volume} {483}},\ \bibinfo {pages} {302} (\bibinfo {year}
  {2012})}\BibitemShut {NoStop}%
\bibitem [{\citenamefont {Uehlinger}\ \emph {et~al.}(2013)\citenamefont
  {Uehlinger}, \citenamefont {Jotzu}, \citenamefont {Messer}, \citenamefont
  {Greif}, \citenamefont {Hofstetter}, \citenamefont {Bissbort},\ and\
  \citenamefont {Esslinger}}]{Uehlinger2013}%
  \BibitemOpen
  \bibfield  {author} {\bibinfo {author} {\bibfnamefont {T.}~\bibnamefont
  {Uehlinger}}, \bibinfo {author} {\bibfnamefont {G.}~\bibnamefont {Jotzu}},
  \bibinfo {author} {\bibfnamefont {M.}~\bibnamefont {Messer}}, \bibinfo
  {author} {\bibfnamefont {D.}~\bibnamefont {Greif}}, \bibinfo {author}
  {\bibfnamefont {W.}~\bibnamefont {Hofstetter}}, \bibinfo {author}
  {\bibfnamefont {U.}~\bibnamefont {Bissbort}}, \ and\ \bibinfo {author}
  {\bibfnamefont {T.}~\bibnamefont {Esslinger}},\ }\href {\doibase 10.1103/PhysRevLett.111.185307} {\bibfield  {journal} {\bibinfo  {journal}
  {Physical Review Letters}\ }\textbf {\bibinfo {volume} {111}},\ \bibinfo
  {pages} {185307} (\bibinfo {year} {2013})}\BibitemShut {NoStop}%
\bibitem [{\citenamefont {Greif}\ \emph {et~al.}(2013)\citenamefont {Greif},
  \citenamefont {Uehlinger}, \citenamefont {Jotzu}, \citenamefont {Tarruell},\
  and\ \citenamefont {Esslinger}}]{Greif2013}%
  \BibitemOpen
  \bibfield  {author} {\bibinfo {author} {\bibfnamefont {D.}~\bibnamefont
  {Greif}}, \bibinfo {author} {\bibfnamefont {T.}~\bibnamefont {Uehlinger}},
  \bibinfo {author} {\bibfnamefont {G.}~\bibnamefont {Jotzu}}, \bibinfo
  {author} {\bibfnamefont {L.}~\bibnamefont {Tarruell}}, \ and\ \bibinfo
  {author} {\bibfnamefont {T.}~\bibnamefont {Esslinger}},\ }\href {\doibase 10.1126/science.1236362} {\bibfield  {journal} {\bibinfo  {journal}
  {Science}\ }\textbf {\bibinfo {volume} {340}},\ \bibinfo {pages} {1307}
  (\bibinfo {year} {2013})}\BibitemShut {NoStop}%
\bibitem [{\citenamefont {Joerdens}\ \emph {et~al.}(2008)\citenamefont
  {Strohmaier}, \citenamefont {Guenther}, \citenamefont {Moritz},\ and\
  \citenamefont {Esslinger}}]{Joerdens2008}%
  \BibitemOpen
  \bibfield  {author} {\bibinfo {author} {\bibfnamefont {R.}~\bibnamefont
  {Joerdens}}, \bibinfo {author} {\bibfnamefont {N.}~\bibnamefont {Strohmaier}},
  \bibinfo {author} {\bibfnamefont {K.}~\bibnamefont {Guenther}}, \bibinfo {author} 
  {\bibfnamefont {H.}~\bibnamefont {Moritz}}, \ and\ \bibinfo {author}
  {\bibfnamefont {T.}~\bibnamefont {Esslinger}},\ }\href {\doibase 10.1038/nature07244} {\bibfield  {journal} {\bibinfo  {journal} {Nature}\
  }\textbf {\bibinfo {volume} {455}},\ \bibinfo {pages} {204} (\bibinfo {year}
  {2008})}\BibitemShut {NoStop}%
\bibitem [{\citenamefont {Scarola}\ \emph {et~al.}(2009)\citenamefont
  {Scarola}, \citenamefont {Pollet}, \citenamefont {Oitmaa},\ and\
  \citenamefont {Troyer}}]{Scarola2009}%
  \BibitemOpen
  \bibfield  {author} {\bibinfo {author} {\bibfnamefont {V.}~\bibnamefont
  {Scarola}}, \bibinfo {author} {\bibfnamefont {L.}~\bibnamefont {Pollet}},
  \bibinfo {author} {\bibfnamefont {J.}~\bibnamefont {Oitmaa}}, \ and\ \bibinfo {author}
  {\bibfnamefont {M.}~\bibnamefont {Troyer}},\ }\href {\doibase 10.1103/PhysRevLett.102.135302} {\bibfield  {journal} {\bibinfo  {journal}
  {Physical Review Letters}\ }\textbf {\bibinfo {volume} {102}},\ \bibinfo
  {pages} {135302} (\bibinfo {year} {2009})}\BibitemShut {NoStop}%
\end{thebibliography}
\end{document}